\documentclass[10pt,aps,pre,twocolumn,superscriptaddress]{revtex4-2}
\usepackage[linkcolor = blue, citecolor = blue, urlcolor = blue, colorlinks = true]{hyperref}
\usepackage[usenames,dvipsnames,x11names]{xcolor}
\usepackage[normalem]{ulem}
\usepackage{graphicx}
\usepackage{listings}
\usepackage{stmaryrd}
\usepackage{amssymb}
\usepackage{amsmath}
\usepackage{gensymb}
\usepackage{float}
\usepackage{bbold}
\usepackage{bm}
\definecolor{crimson}{rgb}{0.75686,0,0.262745}
\definecolor{saphire}{rgb}{0.0,0.196,0.372549}
\definecolor{plum}{rgb}{0.50588,0.007843,0.3843137}

\begin{document}
\title{Avalanches in 
 glassy systems of active particles with finite persistence}
 \title{Avalanches in finite persistent dense disordered active matter}
\title{Avalanches in active glasses with finite persistence}
\author{Roland Wiese}
\affiliation{Computing and Understanding Collective Action (CUCA) Lab, Departament de F\'isica de la Materia Condensada, Universitat de Barcelona, Mart\'i i Franqu\`es 1, E08028 Barcelona, Spain}
\affiliation{UBICS University of Barcelona Institute of Complex Systems , Mart\'i i Franqu\`es 1, E08028 Barcelona, Spain}
\author{Ezequiel Ferrero}
\affiliation{Instituto de Nanociencia y Nanotecnolog\'ia, CNEA–CONICET,
Centro At\'omico Bariloche, (R8402AGP) San Carlos de Bariloche, R\'io Negro, Argentina}
\author{Demian Levis}
\affiliation{Computing and Understanding Collective Action (CUCA) Lab, Departament de F\'isica de la Materia Condensada, Universitat de Barcelona, Mart\'i i Franqu\`es 1, E08028 Barcelona, Spain}
\affiliation{UBICS University of Barcelona Institute of Complex Systems , Mart\'i i Franqu\`es 1, E08028 Barcelona, Spain}

\begin{abstract}
We numerically investigate the statistics of avalanches in glassy systems of active particles with finite persistence, with and without an externally applied shear. 
In departing from the infinite-persistence limit and exploring the interplay of internal activity and external driving, we uncover when and why active and passive systems display similar avalanche statistics and where these analogies fail. 
We find that power-law distributed stress drops emerge only when activity builds long-enough correlations, controlled by the persistence length, with exponents that vary from the purely strain-driven case, to the purely activity-driven case, in a smooth fashion. 
%
The local structure and scaling 
of avalanches of plastic rearrangements remains universal across both limit cases, supporting an interpretation of activity as increasing the typical size of the regions involved in a given avalanche. 
Our results bridge quasistatic shear strain and finite-persistence active yielding, 
showing that avalanches driven by self-propulsion retain the characteristic 
fingerprints of long-range stress propagation. 
\end{abstract}

\maketitle

\section{Introduction}





At high densities, when packing effects become predominant, particulate  matter 
is constrained to evolve collectively, involving structures that exceed by far 
the size of a  particle. 
Disordered or amorphous solids composed of self-propelled units  exhibit 
interesting collective dynamics, triggered by the competition between 
crowding and active driving forces acting at the single particle 
scale~\cite{henkes2011active,ni2013pushing-the-glass-transition, berthier2013non, berthier2014nonequilibrium, mandal2016active,ding2017study,liluashvili2017mode, janssen2019active, berthier2019glassy, henkes2020dense, paoluzzi2022motility, keta2022disordered, karmakar2024perspective-active-glassy, sollich2025elastoplastic-model,GoswamiNP2025}. 
Assemblies of cells or self-driven colloidal suspensions constitute another instance of such dense disordered active systems, displaying glassy-like 
dynamics~\cite{berthier2013non, berthier2014nonequilibrium}. 
A key question to understand dense active matter is how to relate the 
collective dynamics induced by self-propulsion to those observed in 
externally driven systems: 
which distinct self-organization mechanisms, if any, arise uniquely from 
internal activity?
In order to characterize the dynamical properties of active amorphous 
materials one would also like to understand how they respond to external stimuli. 
In the following, we address these questions by investigating a 
model active glass composed of self-propelled particles under shear strain. 
To do so, we investigate the statistics of avalanches triggered by both 
a global drive (strain) and a local one (self-propulsion) separately, as well as the mixed case where both local and global drives are at play.   


Intermittent dynamics and avalanches have been widely reported
in dense particle systems  under external deformation.
These have mostly been studied in the steady state under athermal 
conditions and for very slow or quasistatic driving~\cite{maloney2004subextensive, maloney2006amorphous, robbins2012avalanches, salerno2013effect-of-inertia, LiuPRL2016, dahmen2017scaling, RuscherTL2021, clemmer2021criticality-II, SaitohFP2025}. 
Reports coincide that in this limit amorphous systems undergo a punctuated
evolution of alternating stress-loading and stress-drop stages, with a broad 
distribution of sizes for the so-called `avalanches', correlated sequences 
of plastic events that intermittently release the energy after loading.
The distributions of avalanche size $S$ follow a power-law of the kind
$p(S) \propto S^{-\tau} f(S/S_c)$, with $f(x)$ a rapidly decaying function,  
$S_c$ a cutoff (typically related to a finite system size) and
$S$ defined from the system volume and the related stress drop 
$S\equiv L^d \Delta\sigma$ in $d$ dimensions.

Although some degree of universality is expected for the exponent $\tau$ 
characterizing the avalanche size distributions, from its scaling relations 
with other critical exponents and the theories of driven phase transitions, 
the truth is that one finds a broad range of reported values in the literature. 
For instance, in two dimensions: 
$\tau\simeq 1.2$~\cite{robbins2012avalanches, salerno2013effect-of-inertia}, 
$\tau \simeq 1.28$~\cite{LiuPRL2016}, 
$\tau\simeq 1.5$~\cite{dahmen2017scaling, OyamaPRE2021}, 
$\tau\simeq 0.98$~\cite{RuscherTL2021},
$\tau\simeq 1.3$~\cite{clemmer2021criticality-II},
$\tau\simeq 1.1$~\cite{SaitohFP2025}.
Such discrepancies do not seem to arise from the different kinds of particle
models (be they Lennard-Jones, Kob-Andersen or other interparticle 
potentials typically used in the field) but rather from the dynamical protocols employed. 
Different ways of defining athermal quasistatic (AQS) 
protocols exist and they seem to be determinant in  the outcome of
atomistic models regarding avalanches; not to mention if inertial 
effects are also taken into account, the way systems are overdamped
is also crucial~\cite{salerno2013effect-of-inertia}.
Of course, density or packing-fraction  also plays a role, since 
it affects the efficiency of the relaxation algorithm implemented
in the AQS protocol.

AQS protocols have recently been extended to active matter systems, which means 
that the energy input is not provided by an external load, but originates from 
the self-motion of the constituents of the system themselves. 
In particular, the focus has recently been put on infinite persistence 
models~\cite{morse2021link-active-matter-sheared-granular, keta2023intermittent-relaxation, villarroel2024avalanche-properties}, but the literature also covers
run-and-tumble systems~\cite{ReichhardtNJP2018} and active Lennard-Jones 
glassy systems~\cite{GoswamiNP2025} with finite persistence.
Again, the spread of the avalanche size distribution's critical exponent is
evidenced:
$\tau\simeq 1.0$~\cite{morse2021link-active-matter-sheared-granular},
$\tau\simeq 0.7$~\cite{keta2023intermittent-relaxation},
$\tau\simeq 1.14$~\cite{villarroel2024avalanche-properties},
$\tau\simeq 1.46$~\cite{ReichhardtNJP2018}.
Nevertheless, one observation calling for attention is that 
Ref.~\cite{villarroel2024avalanche-properties}
finds identical avalanche exponents for purely strained
(non-active) and purely active (non-deformed) systems. 
This is somehow surprising, since one doesn't expect \textit{a priori} the
same stick-slip behavior for those very distinct kinds of deformation 
or energy input (external and global\textit{ vs. }internal and local). 
Moreover, since the model's rheology (in particular the flow curves of 
stress $\sigma$ as a function of the strain rate $\dot{\gamma}$) was reported to change considerably, 
giving different Herschel-Buckley exponents~\cite{villarroel2021critical-yielding-rheology, WiesePRL2023}.
This suggests  there might be something in the quasistatic limit (i.e., on the critical point)
that makes the avalanche statistics presumably independent on the type of deformation.
Motivated by all these recent findings in the field, and by the last remark in 
particular, we propose to study avalanche dynamics in a dense mixture of
active Brownian particles (ABP), whose phase diagram in three dimensions has been 
recently characterized~\cite{WiesePRL2023}.
In our analysis, we don't limit ourselves to purely strained or purely active
systems, but study the hybrid situation in which the system is simultaneously 
driven by an external load and an internal activity self-propelling the particles 
at a finite persistence time. 
We go from one limit to the other, applying activity and strain-rate simultaneously,
at intermediate parameter values.
%
Furthermore, we go beyond recent studies, departing from the limiting case
of infinite persistence, which, as we show, oversimplifies the complete 
phenomenology of a much more physically rich and experimentally relevant case of
finite persistence.
%


By studying not only the avalanche statistics for different types of driving, persistence times
and system sizes, but also a `participation'  and other correlation-related
observables, we attempt an explanation on why the scale-free avalanche statistics seems
to be so robust (for a given protocol) when changing from external shear to self-activity.
Apparently, the self-activity introduces a growing length scale to the system, as both the self-propulsion velocity and persistence time  
are increased.
This leads to the rearrangement of larger and larger regions causing the
avalanches, although the related stress drops do not increase much in magnitude. 
`Zooming-out' an active system,  
one is left roughly with the same 
picture as in the non-active, purely externally driven system, suggesting that the size of elementary plastic rearrangements increases with activity. 
We rationalize our findings by comparing the behavior of this emerging length scale 
with the more extreme case of active matter in which Motility-Induced Phase Separation (MIPS)~\cite{cates2015motility} steps in. It has been argued that MIPS can be understood, in an approximated two-body problem, as a long-wave length instability of the  homogeneous state triggered by an effective inter-particle attraction emerging from activity~\cite{bialke2013microscopic, farage2015effective, sese2021phase}. With this in mind, we also compare our results with the behavior of 
passive systems that include cohesive interactions~\cite{SaitohFP2025}. 

Our analysis also reveals that a recently proposed scalar 
observable for active matter systems, the so called `random stress'~\cite{morse2021link-active-matter-sheared-granular, agoritsas2021mean-field, villarroel2021critical-yielding-rheology, villarroel2024avalanche-properties} - which in the infinite persistence 
limit has been proved mathematically to be relevant in mean-field systems - 
is not a good indicator of the avalanche activity in the finite-persistence case.

\section{Results}

\begin{figure}[t!]
\centering
\includegraphics[width=0.9\columnwidth]{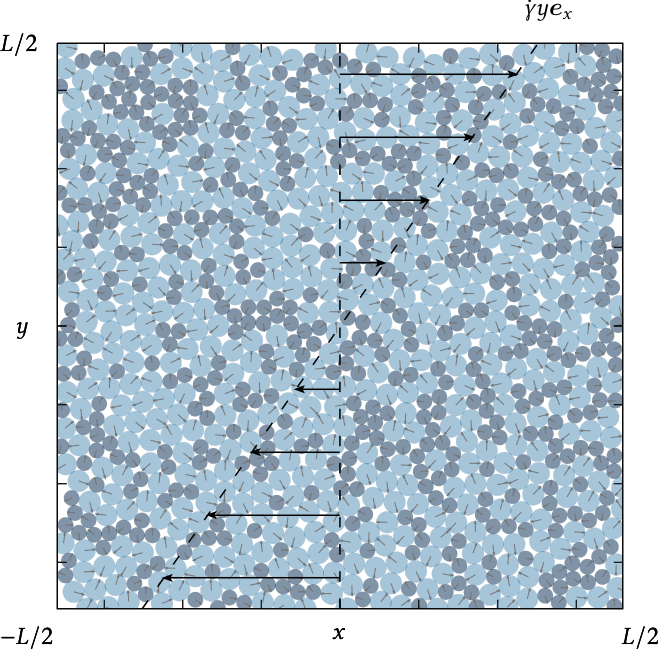}
\caption{
Configuration snapshot of a system composed by $N=10^3$ particles at packing 
fraction $\phi=0.9$ in a $L\times L$ periodic box.
Particles of diameter $d_s=1$ are shown in dark blue while those larger,
of diameter $d_b=1.4$, are in lighter blue.
The arrow inside each disk show its intrinsic self-propulsion direction.
A linear shear profile $\dot{\gamma}y\bm e_x$ is globally applied on 
the system at a fixed strain-rate $\dot{\gamma}$.
}
\label{fig:shear-profile}
\end{figure} 

We perform Brownian dynamics simulations of a 50:50 bidisperse mixture of $N$ 
Active Brownian (ABP) soft disks in a  $L\times L$ box, with a diameter 
ratio of $d_b/d_s=1.4$ between the big and small particles, all of which self-propel 
at a constant swim speed $v_0$ (see Methods section~\ref{sec:methods}). 
To study the system under shear deformation, we impose a linear velocity profile 
along the direction $\hat{\bm x}$ with a shear-strain-rate $\dot{\gamma}$ by imposing Lees-Edwards boundary conditions~\cite{lees1972computer-study-of-transport-processes, allen-tildesley2017computer-simulation-of-liquids}.
Individual particles perform a persistent random walk with a persistence 
time $\tau_p=1/D_r$, $D_r$ being the rotational diffusivity.
The strength of the activity is quantified by the P\'eclet number 
$\mathrm{Pe}=v_0/(d_sD_r)$, that can be written in terms of the persistence 
length $\ell_p=v_0\tau_p$ as $\mbox{Pe}=\ell_p/d_s$.
Figure~\ref{fig:shear-profile} shows a typical snapshot of a system with 
$N=10^3$ particles at a packing fraction $\phi=\pi N(d_s+d_b)^2/(8L^2)=0.9$. 
Results shown in the following are expressed using the small particle 
diameter $d_s$ as the unit of length, the stiffness of the interparticle 
harmonic potential $\epsilon$ as the unit of energy and $d_s^2/(\mu\epsilon)$ 
as the unit of time. 
In the following, we consider a range of P{\'e}clet numbers $\mathrm{Pe}\leq 30$, 
below the MIPS critical point. 
The main control parameters in our study are $\mathrm{Pe} $ and  $\dot{\gamma}$, 
which allow to smoothly tune between regimes  dominated either by activity or 
strain, respectively.

\subsection{Stress drops}

Avalanches of plastic events are unequivocally defined only in 
quasistatic deformations.
Consider starting from a densely jammed system and deforming it very slowly 
(be it externally and globally or internally and locally)
until a first non-affine deformation - a plastic rearrangement - 
takes place. 
At that point one stops the driving and lets the system evolve 
in a cascade of correlated rearrangements caused by the first and only 
driving-triggered one. We can then call `avalanche'  the sequence of such 
consecutive correlated plastic events.
The avalanche stops when no further movement is observed, and one starts deforming 
again quasistatically to trigger the next one. 
These two steps of deformation and energy minimization constitute the so-called athermal quasi-static (AQS) protocol.
After a long sequence of such avalanches, the system reaches a steady state where
one can characterize distributions of avalanche sizes, durations, shapes, 
etc.~\cite{maloney2006amorphous, NicolasRMP2018, robbins2012avalanches}. 

When the deformation is not quasi-static (as in most real cases), the best
thing one can do is to set a criterion to define `avalanches' within a certain tolerance
or threshold uncertainty.
The most common procedure is to analyze stress-strain (or stress-time) series,
recording a scalar stress variable in time as the deformation is applied, and define `avalanches' 
as the stress drop intervals.
The rationale is that, as we load the system, the stress goes up until a plastic rearrangement 
is triggered which in turn can trigger others and that collective phenomenon causes 
a global stress-drop.
Stress drops, denoted $\Delta\sigma$,  can be defined by analyzing the negative intervals of the stress-time derivative signal, commonly shifted by a threshold to discard 
signal noise in experiments~\cite{VivesPRE2016, AlavaPRL2016, MunozPRE2019}, 
see Fig.~\ref{fig:stress-time-series} for a graphic definition.
Then the avalanche size $S$ is simply defined as an extensive quantity 
multiplying the stress drop by the system size
\begin{equation}
    S \equiv L^d \Delta\sigma
\end{equation}
in $d$ dimensions.

Stress drops are detected by monitoring sign changes of the 
slope of the stress $\sigma(t)$, meaning that they start at 
time $t_i$ when the slope becomes negative, determined by the 
two conditions 

\begin{align}
\begin{split}
    \sigma(t_i + \Delta t) - \sigma(t_i)&<0\,,\\
    \sigma(t_i) - \sigma(t_i-\Delta t) &>0\,.
\end{split}
\label{eq:time-criterion}
\end{align}

Their end-time $t_f$ is detected likewise for the slope becoming  positive, 
with the inequalities in the criterion reversed.
Our time series were sampled with a time step $\Delta t=1$,  i.e. at every 
hundredth integration step $dt=0.01$
\footnote{We note that both the time step 
$\Delta t$ and the threshold used to define avalanches affect numerical details of the distributions. Variations of the latter mainly shift the power-law regimes discussed in the next section, without altering the scaling exponent.
For consistency, we use a zero threshold, to avoid parameter optimization, while keeping the results stable.
}. 
 In order to ensure that a steady state is reached, we discard a transient 
strain, $\dot{\gamma}t \sim 1$, for the sheared system, and a similar 
accumulated strain $t/\tau_p \sim 1$ for the purely-active and combined 
sheared-active systems (see \cite{SM} for time series of the random stress?).

\subsection{Distribution of avalanche sizes }

\begin{figure}[t!]
\centering
\includegraphics[width=\columnwidth]{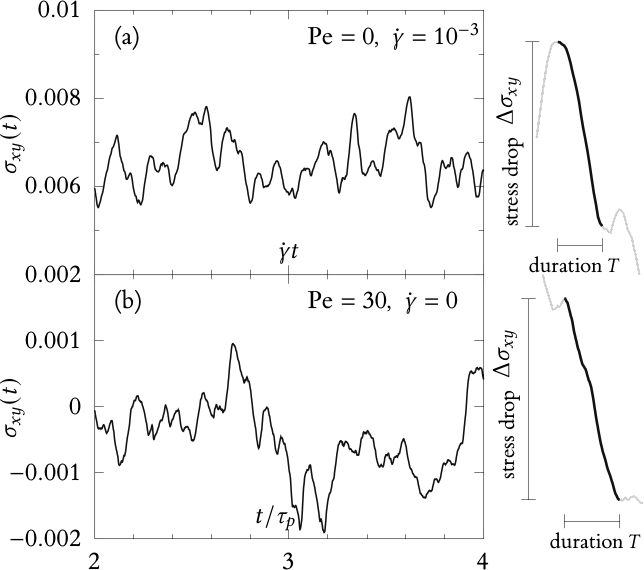}
\caption{
Example time series of the Irving-Kirkwood stress $\sigma_{xy}(t)$ in the passive sheared system in (a) and the active unsheared system in (b), for $N=10^{3}$ and $\phi=0.9$ in both cases.
On the right, the identification of a single avalanche via sign changes in the slope of $\sigma_{xy}(t)$ is illustrated, as described in the text. 
}
\label{fig:stress-time-series}
\end{figure} 

We measure the off-diagonal component of the Irving-Kirkwood (IK) stress 
tensor (see Methods). 
In the steady state, it fluctuates around an average value. 
In the driven case, this value is finite and depends on the strain rate 
through a Herschel-Bulkley law $\sigma=\sigma_y+A\dot{\gamma}^n$, where $\sigma_y$ 
is the yield stress and $n$ a critical exponent~\footnote{The critical aspect of 
the yielding transition is better seen when considering the strain rate as 
an order parameter and the stress as the control parameter~\cite{FerreroSM2019, NicolasRMP2018}}. 
In steady conditions, one defines the shear viscosity from the IK stress 
as $\eta(\dot{\gamma}) = \langle\sigma_{xy}\rangle/\dot{\gamma}$. 
On the other hand, for the purely active system $\sigma_{xy}$ fluctuates 
around zero. 
To illustrate these points, we show 
examples of time series $\sigma_{xy}(t)$  in Fig.~\ref{fig:stress-time-series}, plotting the IK stress over 
comparable time windows for a purely sheared and a purely active case.
Beyond their average value, the fluctuations in stress, and in particular the drops,
look similar and are broadly distributed, as will be discussed in the following.

\begin{figure}[]
\centering
\includegraphics[width=\columnwidth]{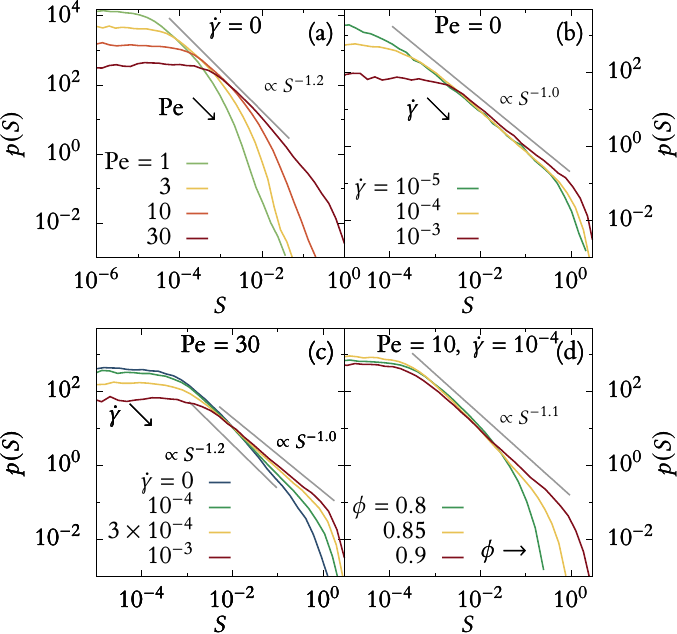}
\caption{
The effect of activity and shearing on avalanche size distributions $p(S)$ of the IK stress $\sigma_{xy}$ for $N=10^3$ particles at $\phi=0.9$ (except in (d), as indicated).
(a) Increasing the P\'eclet number at zero shear suppresses small avalanches and boosts larger ones, until a power-law with exponent $\tau\approx1.2$ emerges at $\mathrm{Pe}\approx30$.
(b) Increasing the shear rate $\dot{\gamma}$ also suppresses small avalanches, but the distribution decays more slowly with $\tau\approx1.0$ and the tails of $p(S)$ are not as strongly affected.
(c) Applying both active and sheared forcing at $\mathrm{Pe}=30$, the activity dominates for $\dot{\gamma}\leq 10^{-4}$, reflected in the faster decay of $p(S)$ with exponent $\tau\approx1.2$.
As $\dot{\gamma}$ increases, the limiting form of decay with $\tau\approx1.0$ is reached at $\dot{\gamma}=10^{-3}$ for dominating shear.
(d) The dependence on $\phi$ for high densities $\phi\geq0.8$ is relatively weak and does not strongly affect the power-law scaling, but reduces its range. 
} 
\label{fig:av-overview}
\end{figure}

We now turn to the statistics of avalanche sizes, which is summarized in Fig. \ref{fig:av-overview} for a generously explored range of control parameters $\mathrm{Pe}$, $\dot{\gamma}$, $\phi$.
%
In general, power-law distributions of the type $p(S)\propto S^{-\tau}$ with a rapid exponential cutoff are observed.
Distributions of avalanche sizes in the purely active system are shown in Fig.~\ref{fig:av-overview}(a) and an increasing $\mathrm{Pe}$ is observed to produce larger avalanches.
For our largest activity, $\mathrm{Pe}=30$, the tail of $p(S)$ 
becomes broadest, expanding the parameter range where we observe a power-law decay  with an exponent that we estimate to be $\tau\approx1.2$.
%

Fig.~\ref{fig:av-overview}(b) displays the other extreme case of
non-active and purely driven systems ($\mathrm{Pe}=0$, $\dot{\gamma}>0$).
Increasing the strain rate $\dot{\gamma}$ suppresses small avalanches and 
slightly induces the occurrence of larger ones, shifting the cutoff to larger $S$.
Yet, all these shear-induced avalanche size distributions consistently 
display an exponent $\tau\!\approx\!1$.

The mixed effect of applying external shear on an active system is 
shown in Fig.~\ref{fig:av-overview}(c) for $\mathrm{Pe}=30$ and $\dot{\gamma}\in[0,10^{-3}]$.
Interestingly, while for shear rates $\dot{\gamma}\lesssim 10^{-4}$ the statistics 
seems to be governed by the ``active exponent'' $\tau\approx1.2$, $\tau$ 
decreases as the strain rate increases, and the external shear begins to dominate 
the avalanche size statistics for $\dot{\gamma} \gtrsim 10^{-3}$. 
Besides the numerical values of the exponents that one could extract 
from such data, one clearly finds that $p(S)$ decays faster as the activity 
becomes dominant. 

Finally, Fig.~\ref{fig:av-overview}(d) illustrates the dependence of avalanche size on the density $\phi$ for $\mathrm{Pe}=10$ and $\dot{\gamma}=10^{-4}$.
Decreasing $\phi$ from its highest value $\phi=0.9$ 
(the one used in Figs.~\ref{fig:av-overview}(a-c)),
we observe a narrowing of the power-law region, with the upper cutoff moving 
to smaller values of $S$. 
In this high density regime we focus on, the system is close to its dynamic arrest. Indeed, the system exhibits a transition from a fluid to an arrested (glassy) state at packing fractions that range from around $\phi\approx0.8$ in the passive case, up to $\phi\approx 0.87$ for $\mathrm{Pe}=10$ (see \cite{SM} for details on the dynamical arrest). 
As shown in Fig.~\ref{fig:av-overview}(d), the range where a power-law decay of $p(S)$ is observed extends as $\phi$ increases while keeping the exponent roughly unchanged, 
leaving the idea that $\tau$ is controlled by the interplay between 
$\dot{\gamma}$ and $\mathrm{Pe}$ rather than by the density.
Note that, however,  
 it is not crucial to be in the arrested regime to observe broad avanche distributions. 
  ~\cite{robbins2012avalanches,NicolasRMP2018,RuscherTL2021},
Activity  has a mixed effect: it shifts the glass transition to higher packing 
fractions while it builds larger stress drops. 

In conclusion, both external shear strain and internal activity have the 
ability to lead the system to a self-tuned `critical' state where avalanche
statistics show fat-tailed power-law distributions for the avalanche size $S$.
Increasing the driving, be it external or internal, tends to populate the distributions at larger values of $S$.
Nevertheless, the main parameters controlling the upper cutoff of the 
distributions are the system's density $\phi$ and the system size $L$, the 
impact of which is analyzed in the next section.
%
With respect to the values that we estimate for the $\tau$ exponent, 
they can be considered to be compatible within error bars with the ones 
reported in the works by 
Morse \textit{et al.}~\cite{morse2021link-active-matter-sheared-granular} 
($\tau\simeq 1.0$) and Villarroel and 
D\"uring~\cite{villarroel2024avalanche-properties} ($\tau\simeq 1.14$), 
which are both computed under AQS deformation in infinite persistence 
systems.
Yet, the main difference in our data is a systematic observation of
a small but appreciable exponent change when one switches between 
a purely activity-driven ($\tau\sim 1.2$) and a purely 
shear-driven system ($\tau\sim 1.0$), in contrast 
with the observations in Ref.~\cite{villarroel2024avalanche-properties}.
Compared to the literature on non-active driven systems, 
the exponent of our avalanche size distributions for $\mathrm{Pe}=0$
seems to better agree with the one reported  in Lennard-Jones binary mixtures 
($\tau\sim 1.0$ vs. $\tau\sim 0.98$) ~\cite{RuscherTL2021}.
All other studies reported larger exponents~\cite{robbins2012avalanches, salerno2013effect-of-inertia,LiuPRL2016,dahmen2017scaling, OyamaPRE2021,
clemmer2021criticality-II,SaitohFP2025}.

Activity introduces both a persistence time and length. 
For a given persistence time, each $\mathrm{Pe}$ has an 
associated $\ell_p$, that we argue controls the size of 
the plastic rearrangements. 
All observations are consistent with plastic rearrangements becoming larger both with increasing persistence time and persistence length. The role of persistence in the yielding of dense active systems has been recently discussed in Ref.~\cite{GoswamiNP2025}. Beyond a “thermal limit’’ at small persistence times and an initial non-monotonic behaviour, highly persistent active forces increasingly work to unpin the system from jammed, kinetically arrested states. As persistence grows and particles lose the flexibility to explore alternative pathways, they must effectively “break through’’ their cages, requiring progressively larger active forces.
In line with this picture, we find that the associated stress drops also grow (on average) with increasing persistence time. In the steady state, where one expects the stress to fluctuate around a plateau value, if the system must build up to higher stresses to escape arrested configurations, the resulting stress releases are necessarily larger as well, not event by event,  but on average.
%
%
When varying $\mathrm{Pe}$ at a fixed system size $N$, 
as in Fig.~\ref{fig:av-overview}(a),
we are effectively changing the number of events that 
`fit' in the system, and therefore some rescaling 
of $S$ is expected to hold.
Proposing a scaling of the form $S \to S/\ell_p^b$, the $p(S)$
curves in Fig.~\ref{fig:av-overview}~(a) collapse fairly well 
with $b=1$, see the SM~\cite{SM}.
%
%
%

%

\begin{figure}[t!]
    \centering
    \includegraphics[width=\linewidth]{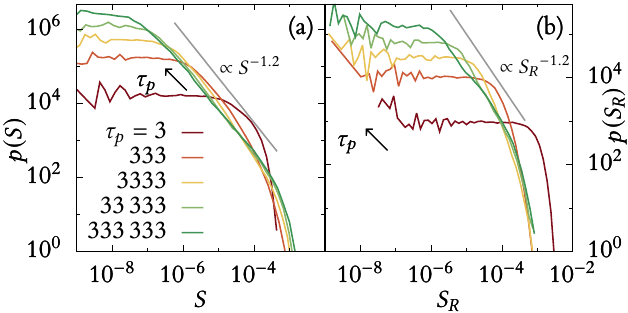}
    \caption{The effect of changing the persistence time $\tau_p$ on purely active avalanches at $\dot{\gamma}=0$ is shown for constant  self-propulsion velocity $v_0=0.009$ (corresponding to $\mathrm{Pe}=30$ for $\tau_p=3333$).
    IK stress drops $S$ in (a) develop a power-law scaling with $\tau\approx1.2$.
    The drops in the random stress $S_R$ in (b) similarly show a power-law at the highest $\tau_p$, although over a smaller range in $S_R$. 
    }
    \label{fig:av-random}
\end{figure}

We complement the analysis of avalanche statistics computed from the IK stress 
with measurements of the random stress,
\begin{equation}
    \sigma_R(t) = - \frac{v_0}{L^2}\sum_{i=1}^N\bm F_i(t)\cdot\bm n_i(t)\,,
\end{equation}
defined as an anticorrelation between interactions $\bm F_i$ and active forces $v_0\bm n_i$.
The random stress $\sigma_R$ was introduced in the context of studies of active yielding
in systems with infinitely persistent active particles~\cite{morse2021link-active-matter-sheared-granular, villarroel2024avalanche-properties}.
It has been reported that in the limit of quasistatic driving by active forces it yields the same avalanche statistics as the IK stress $\sigma_{xy}$ under external shear strain~\cite{villarroel2024avalanche-properties}.
Moreover, in the mean field under athermal quasistatic conditions, $\sigma_R$ is 
argued to be formally equivalent to a shear stress, see~\cite{agoritsas2021mean-field, morse2021link-active-matter-sheared-granular}. 
In the steady state, we find $\langle\sigma_R\rangle\propto\mathrm{Pe}$.
%
From the stress drops of the random stress $\sigma_R$, one can also define
a distributions $p(S_R)$ of avalanche sizes $S_R$.
Examples are shown in Fig.~\ref{fig:av-random}(b) and compared to the 
$p(S)$ distributions in the next section. 
%
%
The signal of the random stress is not as clean as the for IK stress, especially for short persistence times.
Consequently, the avalanche distributions for small and moderate
$\tau_p$ look quite different between $p(S)$ and $p(S_R)$~\footnote{Even though we have tried to make the analysis of both stress signals 
compatible, e.g., by detecting the avalanche time windows in $\sigma(t)$
and using the same starting/ending points for both $\sigma(t)$ and $\sigma_R(t)$.
}.
Yet, one expects full consistency (for the avalanche statistics) between 
both stress definitions in the limit of infinite persistence. 
A precursor of this is seen when comparing the shapes of $p(S)$ and $p(S_R)$ in Fig.~\ref{fig:av-random} at the largest $\tau_p$.

\subsection{The role of persistence}

%
The effect of changing the persistence time $\tau_p$ on the avalanche size distributions $p(S)$ and $p(S_R)$ is summarized in Fig.~\ref{fig:av-random}~\footnote{To obtain the results shown in Fig.~\ref{fig:av-random} we have changed 
the persistence time $\tau_p$ while keeping a fixed velocity of 
self-propulsion $v_0=0.009$, which amounts to $\mathrm{Pe}=30$ for $\tau_p=3333$.
This was done to avoid the active force becoming stronger than the steric 
repulsion forces. Otherwise, artifacts could arise in the thermal limit of small
persistence when the two forces become comparable, allowing particles to penetrate 
each other.}.
We observe that for very short persistence (high rotational diffusivity), $\tau_p=3$, the stress drops $S$ and $S_R$ 
are exponentially distributed.
This is expected, since a randomly oriented and rapidly diffusing drive should lead to an effective thermal agitation, producing incoherent fluctuations on top of an average stress.
As $\tau_p$ increases, the active forcing has the chance to induce plastic 
rearrangements, which in turn facilitate others and build up correlations.
We observe a gradual change from an exponential to a
power-law distribution, 
which occurs at much lower $\tau_p$ for avalanches $S$ in the IK stress than for the random stress avalanches $S_R$.
%
Considering the IK results in Fig.~\ref{fig:av-random}(a), a clear power-law regime appears for $\tau_p\geq3333$
(we study a persistence $\tau_p=3333$, corresponding to $D_r=3\times10^{-4}$, for the remainder of this work). 
Since the durations $T$ of the avalanches are at most of the order $\mathcal O(10^2)$,
there is already a clear separation of time scales (see~\cite{SM} for $p(T)$ distributions).
For avalanche statistics purposes at least, 
such a persistence could thus be considered to be close enough to the infinite 
persistence limit.
Focussing on the persistence length, having fixed $v_0=10^{-2}$ we get $\ell_p\sim 10^2$ for $\tau_p\sim 10^4$, which is of the order of the box length $L$. 
In this regime we thus expect for the avalanche size distributions a power-law 
decay with an exponential cutoff due to finite system-size.
However, this is only seen in the curves of $p(S)$, 
while $p(S_R)$ still shows an exponential distribution for $\tau_p=3333$
and a power-law begins to emerge only for ten to hundred times higher persistence times.
This indicates that $S_R$ might be a less ideal choice for an observable 
at finite persistence.
%

Notice that curves for $\tau_p\ge 3333$  show $p(S)\propto S^{-\tau}$ 
with $\tau \approx 1.2$ in Fig.~\ref{fig:av-random}(a), consistent 
with the previously discussed results in Figs.~\ref{fig:av-overview}(a),(c).  
However, an attempt to estimate $\tau$ from the $\tau_p=333$ curve would yield 
a larger exponent over a smaller range of $S$ values
and there seems to be a slight decrease of $\tau$ for increasing 
persistence, which converges within the explored parameter values.
A similar effect is observed for increasing system sizes $N$, 
see Sec.~\ref{sec:fss}.
If we look at the distribution of avalanche sizes computed from the random stress
in Fig.~\ref{fig:av-random}(b), 
a power-law $p(S_R)\propto S_R^{-\tau}$ with $\tau \approx 1.1$ can be argued only 
for the largest persistence. 
this is close to the AQS exponent $\tau\approx1.14$ of
Ref.~\cite{villarroel2024avalanche-properties} measured in the infinite 
persistence limit. 
As the persistence time decreases from this extreme case, the range over which 
one observes a power-law decay shrinks and eventually the distribution 
becomes exponential.

Overall, in a finite persistence scenario
we find the IK stress drops to be a more robust measure to quantify avalanches 
than the `random stress' proposed in Refs.~\cite{morse2021link-active-matter-sheared-granular, agoritsas2021mean-field, villarroel2024avalanche-properties}, 
also in the absence of shear.
It could be the case that such a stress definition for active yielding studies
is relevant only in idealized scenarios when persistence is infinite.

\subsection{Finite size scaling analysis}
\label{sec:fss}

\begin{figure}[t!]
    \centering
    \includegraphics[width=\columnwidth]{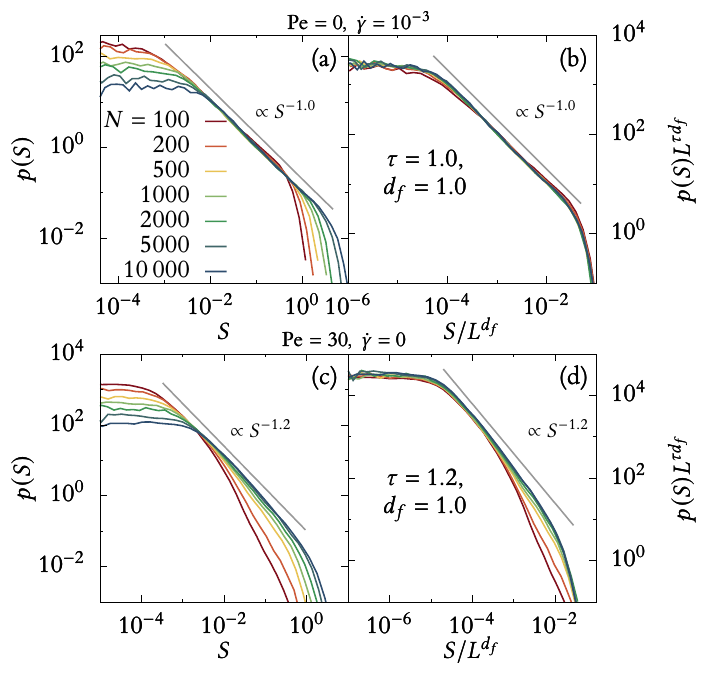}
    \caption{Finite size scaling of the avalanche size distributions $p(S)$ in the passive, sheared system at $\mathrm{Pe}=0$, $\dot{\gamma}=10^{-3}$ in (a),(b) and the active system at $\mathrm{Pe}=30$, $\dot{\gamma}=0$ in (c),(d). 
    }
    \label{fig:av-FSS}
\end{figure}

In an attempt to  better characterize the avalanche  statistics, we perform a finite  
size scaling analysis.
We compute $p(S)$ for 7 different system sizes (particle numbers $N$) ranging 
over two orders of magnitude $N\in[10^2,10^4]$.
We focus on the two very distinct deformation scenarios of pure shearing 
(at $\dot{\gamma}=10^{-3}$, $\mathrm{Pe}=0$) and pure activity 
(at $\dot{\gamma}=0$, $\mathrm{Pe}=30$).
These parameter sets have previously been identified in Fig.~\ref{fig:av-overview}
as avalanche statistics dominated by $\dot{\gamma}$ and $\mathrm{Pe}$, respectively.

Figure~\ref{fig:av-FSS} shows on its left panels the bare $p(S)$ distributions 
for these two cases, for differing system sizes $N$. 
For these distributions the classical dependence on $N$ is observed: 
as system size increases, the cutoff of the power-law in $p(S)$ 
shifts to larger values; at the same time, due to normalization the
height of the initial plateau at small $S$ decreases. 
One expects a scaling law
\begin{equation}
    p(S) = S^{-\tau}f(S/S_c),
\end{equation}
\noindent where $f(x)$ is a rapidly decaying function and $S_c=L^{d_f}$
is a finite system size cutoff.
This scaling defines the fractal dimension $d_f$, 
which we test by plotting $p(S)L^{\tau d_f}$ vs $S/L^{d_f}$ in the panels on the right,
Figs.~\ref{fig:av-FSS}(b)(d).
For the purely sheared case, a very good collapse is achieved for all curves
with $\tau=1.0$ and $d_f=1.0$.
A fractal dimension around $\sim1$ is in agreement with previous reports 
in the literature of avalanches in amorphous 
solids, both in elastoplastic models and molecular 
dynamics~\cite{Talamali2011, LinPNAS2014, NicolasRMP2018, 
LiuPRL2016, FerreroSM2019} and indicates that in 
two-dimensional systems yielding-related avalanches are quasi-1D 
slip lines.
Interestingly, in the active system the same fractal dimension $d_f=1.0$ 
produces a good collapse of the distributions at larger $N$.
The fact of $d_f$ being $\simeq 1$ also in the active case with a randomly 
oriented internal driving is surprising, naively one could have imagined that
the avalanches would be more compact objects.
%
In contrast with the sheared case, we observe that for the value of the P\'eclet number considered in Fig.~\ref{fig:av-FSS}(c),(d), $\mathrm{Pe}=30$, the power-law does not properly evolve until $N>500$.
Any estimation of $\tau$ would therefore vary with $N$ for small 
particle numbers, but a good and stable estimation is found at
$\tau=1.2$ if we restrict ourselves to $N>500$.
Together with $d_f=1.0$ these curves show a good collapse within the expected finite-size scaling law. 

\begin{figure*}[t!]
\centering
\includegraphics[width=\textwidth]{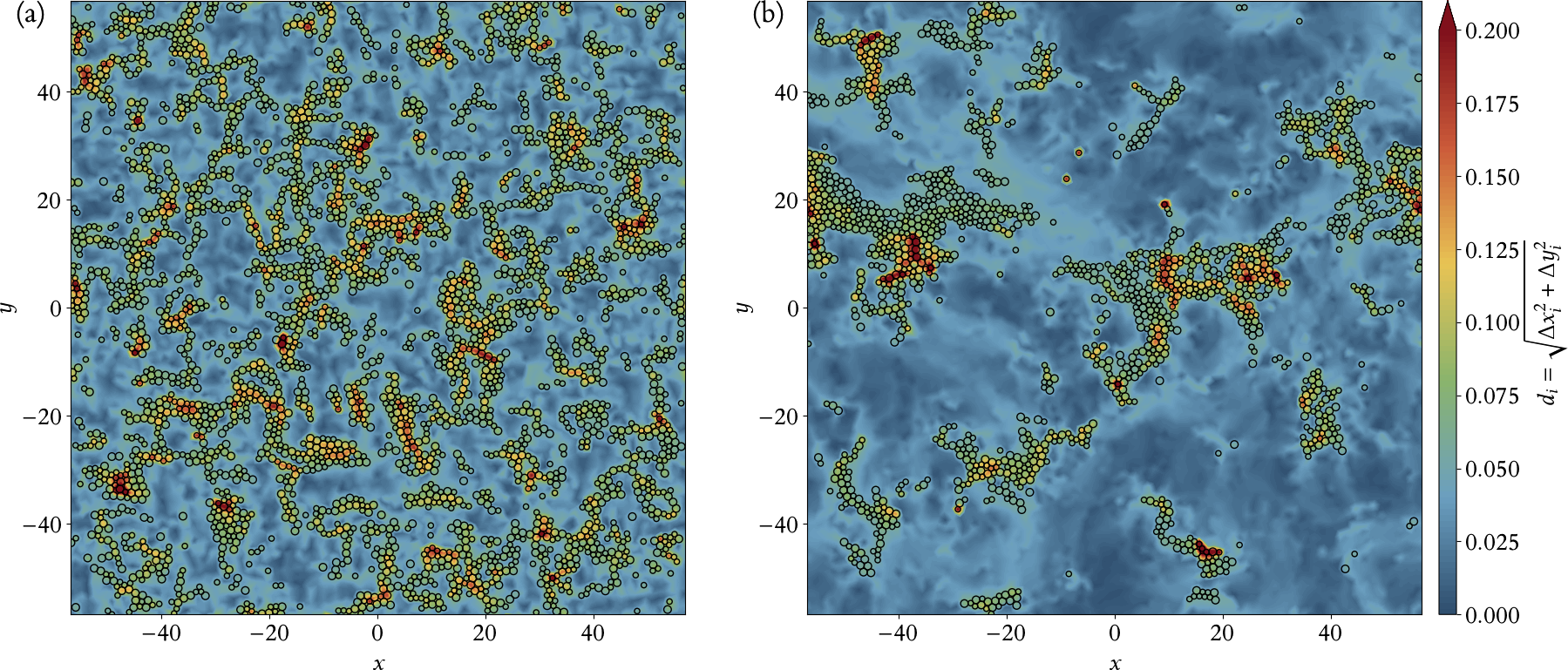}
\caption{
Snapshots of ``mobile'' particles, as defined from the participation eq. \ref{eq:participation}, in a $N=10^4$ system associated to avalanches of size $S\approx 1$ for (a) $\mathrm{Pe}=0$, $\dot\gamma=10^{-3}$ with $\lfloor NP\rfloor=3174$ and (b) $\mathrm{Pe}=30$, $\dot\gamma=0$ with $\lfloor NP\rfloor=1547$.
The background color represents the amplitude of (non-affine) displacements $d_i$ for all particles, with its scale cut off at $d_i=0.2$ for better visibility ($0.2\%$ of the particles displaced with $d_i\in[0.2,0.3]$, also shown in dark red).
The black disk outlines correspond to mobile particles.
The purely active avalanche in (b) shows a smaller number of clusters, while clusters 
in the passive sheared system in (a) are more homogeneously spread out.
The visual difference between the two scenarios is due to the lower number of mobile particles at comparable generated stress in (b), with the structure of clusters being statistically similar (cf. Fig.~\ref{fig:cluster-sizes} below).
}
\label{fig:displacements}
\end{figure*}

As we mentioned earlier, for a given persistence time, each $\mathrm{Pe}$ has an 
associated persistence length $\ell_p$, that seems to control 
the size of the plastic rearrangements, which increases with growing $\ell_p$.
If the system size is comparable to or smaller than this typical size,
the statistics of correlated rearrangements that leads to the power-law in $p(S)$
cannot develop, and the power-law regime is interrupted earlier by a broadly 
distributed (not exponentially fast) cutoff.
We argue that this causes the faster decay observed for 
$N<500$ in Fig.~\ref{fig:av-FSS}(c),(d),
since $\ell_p=d_s\mathrm{Pe}=30$ 
%
is close to the linear size $L=32$ of the 
system with $N=500$ particles. 
Only for $L>\ell_p$ can one appreciate  events beyond the finite size cutoff, leading 
to an algebraic regime. 
As is shown in Fig.~\ref{fig:displacements}, such a  length scale is qualitatively 
compatible with the  typical size of the rearrangements we observe in the 
simulations. 

\subsection{Microstructure of plastic events}


We now turn to the comparison of the spatial distribution and structure of 
avalanches between the classical shear deformation case and the active case. 
%
We define the so-called ``participation'' $P$ of particles during an 
avalanche, defined as the inverse fourth moment of the non-affine displacements~\cite{lemaitre2007avalanche-size} (see Sec.~\ref{sec:methods}).
This quantity has the advantage of not introducing a prescribed, arbitrary 
threshold in the particle displacements, but still providing a measure 
of the number of mobile particles during a given time lapse. 
%
We then proceed to identify the number of particles $\lfloor NP\rfloor$ 
with the largest displacements $d_i$ during an avalanche as the set of 
`mobile particles'.

In order to visualize two `comparable' avalanches (of the same size $S\approx1$),
one triggered by simple shear and the other by self-propulsion, we present in Fig.~\ref{fig:displacements} configuration snapshots of these two systems which 
only show the previously defined mobile particles.
We superimpose the non-affine displacement map across 
the whole system (shown with a color scale). 
Note that the avalanches of size $S\approx 1$ are large and correspond to the tail of the 
sampled distributions (see Fig. \ref{fig:av-FSS}).
%
The passive sheared system on the left and the active system on the right appear 
visually different, with the number of mobile particles in the passive system being roughly twice as large. 
The active system is able to generate a stress drop of the same magnitude $S\approx1$
by involving only half the number of mobile particles.
%
%
%
We therefore expect that higher   particle mobilities will trigger larger 
  stress drops.
This picture agrees with the observation of bigger tails in $p(S)$ for increasing 
P\'eclet numbers, see Fig.~\ref{fig:av-overview}(a).
As $\mathrm{Pe}$ increases, so does the persistence length, building correlations 
at length scales growing with $\sim \ell_p$. 

Further inspecting Fig.~\ref{fig:displacements}, we observe the emergence of mobile clusters in both the passive and active case. 
In the active case, the clusters appear larger, but count fewer mobile particles (in absolute numbers $\lfloor NP\rfloor)$.
Statistically, however, these clusters appear to be scale invariant and therefore 
equivalent by a mere scale transformation, meaning that one can interpret the 
active system as a `zoom-in' on the passive system (in the infinite 
system-size limit).  

\begin{figure}[t!]
\centering
\includegraphics[width=\columnwidth]{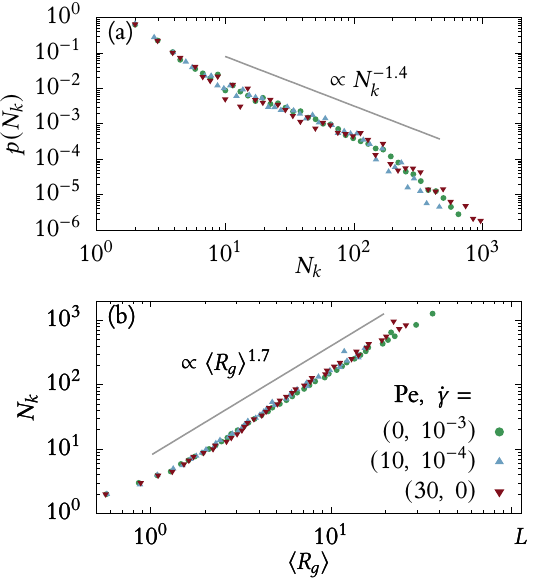}
\caption{
(a) Distribution of cluster sizes $N_k$  
for $\mathrm{Pe}=30$, $\dot{\gamma}=0$ and $\mathrm{Pe}=0$, $\dot{\gamma}=10^{-3}$. A power-law decay $N_k^{-1.4}$ is shown for comparison.  
(b) Scatter plot of the cluster size $N_k$ against their averaged radius of gyration $\langle R_g\rangle$. 
The data shows the same overall scaling for both the passive and active case. 
The results shown in (a) and (b) were obtained for avalanches sizes in the interval $S\in[0.9,1.1]$ in systems of  $N=10^4$ particles.
}\label{fig:cluster-sizes}
\end{figure} 

To provide a quantitative ground for such an interpretation, we analyze the size 
statistics of mobile clusters. 
Figure~\ref{fig:cluster-sizes}(a) shows the distribution of cluster sizes $p(N_k)$,
identified as sets of $N_k$ mobile particles within a distance of 
$r_{\text{cut}}=1.4=d_b$ from each other.
Interestingly, both activity and simple shear deformation lead to the same 
cluster size distribution, that exhibits an algebraic decay 
\begin{equation}\label{eq:PNk}
p(N_k)\sim N_k^{-\alpha}    
\end{equation}
with $\alpha\approx 1.4$, no lower cutoff and an upper cutoff set by the system 
size only.
This means that clusters such as those represented in Fig.~\ref{fig:displacements} 
have the same statistical properties, regardless of whether they appear smaller or bigger,
and can be mapped onto each other by a scale transformation which leaves the distribution $p(N_k)$ invariant.

The relation between the cluster size $N_k$ and its radius of gyration $R_g$ 
defines the cluster's fractal dimension $D_c$ as
\begin{equation}
    N_k\propto R_g^{D_c}\,.
\end{equation}
Figure~\ref{fig:cluster-sizes}(b) depicts a scatter plot of the measured 
radii of gyration \emph{vs.} the corresponding size, or mass, of each cluster. 
The data obtained is compatible with fractal clusters, with a cluster 
fractal dimension $D_c=1.7$, compatible with diffusion limited aggregation~\cite{halsey2000diffusion}. 
Again, we find the same scale-free behavior for passive and active systems. 
Notice that $D_c \neq d_f$; in particular $D_c > d_f$. 
This is similar to what has been observed in classical yielding, 
where for instance $3D$ molecular dynamics of amorphous systems 
under deformation show estimations of 
$D_c\simeq 2$~\cite{GhoshPRL2017, LeishangthemNC2017} 
and $d_f\simeq 1.5$~\cite{LiuPRL2016} or even 
smaller $d_f\simeq 1.15$~\cite{clemmer2021criticality-II}.
While it may look contradictory, since one is talking about 
two `fractal dimensions' of the same object (an avalanche), this 
discrepancy can be understood from the exponents' definitions.
The number of regions (or particles) deforming during an avalanche 
and the avalanche intensity (or total stress drop) are related 
but not necessarily in a linear way 
\footnote{Some regions can deform many times during an avalanche and 
contribute differently to its total `size' $S$ and `duration' $T$.
This is clear in lattice models, were $S \sim T^{d_f/z}$, with $z$ 
the dynamical exponent, but if we count the total `number of activations'
$S_A$, it typically scales with $T$ with another exponent.}.

%
%

Distributions of the participation $p(P)$ for avalanches of different sizes $S$ 
are shown in the Supplemental Material~\cite{SM}.
While they highlight that $P$ is systematically lower in the case of pure activity 
($\mathrm{Pe}=30$, $\dot{\gamma}=0$) as compared to pure shear ($\mathrm{Pe}=0$, 
$\dot{\gamma}=10^{-3}$), it also shows that participation is larger for bigger 
avalanche sizes $S$ in all cases. 
This is consistent with our previous discussion: fewer particles are needed to 
generate a given stress drop as activity is increased. 
Not surprisingly, the case of mixed active and sheared forcing stands in between 
the two extreme cases also in the participation distribution.
Interestingly, the average value of mobile particles 
$\left<NP\right>\equiv\left<\left(\sum_id_i^2\right)^2/\sum_id_i^4\right>$,
observed at a given $S$, systematically increases either with $\dot{\gamma}$ 
or (more gently) with $\mathrm{Pe}$. 

\section{Discussion}

We have simulated and analyzed the statistics of avalanches in glassy 
systems of active particles with finite persistence, with and without 
an additional external shear. 
By departing from the infinite persistence limit (almost exclusively
the case covered in previous literature) and studying the 
combination of internal activity and external shear 
we have been able to elucidate features and reasons for 
the similarity in avalanche statistics observed between passive 
and active systems and for its breakdown.

In brief, the similarity of shearing and activity, observed for 
quenched particle orientations, does not completely carry over 
to finite persistence, although some features remain:
(i) 
In both cases, avalanches of plastic events (quantified by stress drops) 
exhibit power-law distributions if the drive is able to build large enough 
correlations.
In the passive case these are cut off only by the finite size of 
the system.
In the active system, short persistence times can work against 
the build up of correlations.
The avalanche size distributions feature similar exponents, from 
$\tau\approx 1.0$ (driven system, shear-induced) to $\tau\approx 1.2$ 
(active system, activity-induced).
(ii) 
The nature of the avalanches, in terms of the structure of plastic 
events or the spatial distribution of particles involved, 
is the same for active and passive systems. 
The $\alpha$ and $D_c$ exponents characterizing the cluster sizes statistics
are the same in both cases. 
We interpret this scale invariance as rooted in the fact that larger intrinsic 
particle mobilities induce larger (more numerous) clusters of locally jammed 
particles and therefore require larger collective motions to play the role
of the elementary shear-transformation-zones equivalents.
This is compatible with the picture for sticky grains~\cite{SaitohFP2025}, 
where more attractive particles induce larger elementary plastic rearrangements,
providing a way of qualitatively interpreting activity as an effective 
two-body stickiness~\cite{ginot2015nonequilibrium, farage2015effective}. 

At finite persistence, stress drops defined using the so-called 
\textit{random stress}, originally introduced for quenched quasistatic (QS) 
protocols, are less effective statistical indicators of correlated events than 
those obtained from the classical Irving–Kirkwood stress tensor. 
Nevertheless, it provides a natural link between QS approaches and our 
simulations at very large persistence times, thereby connecting previous
results in the infinite-persistence limit with those presented here.

An important remark about what we learn from this work 
is that even in the cases where yielding is driven by 
self-propulsion rather than external load, it continues to be a phenomenon 
characterized by a non-concave (or non-positively defined) long-range 
stress propagator, 
whose fingerprints in the statistics of avalanches are an exponent 
$\tau$ clearly below the depinning mean-filed exponent $\tau=1.5$ and 
a fractal dimension $d_f$ much smaller than the spatial dimension (non-compact objects).
The fact that, given a methodology and a defined dynamical protocol, these exponents 
look also very close in the extreme cases of purely driven and purely active systems
remains a surprising feature, but we attempted to find a rationalization 
to that end.

The basic understanding of deformation of amorphous materials with active 
constituents is not only relevant in cell biophysics~\cite{lenne2022sculpting, mongera2018fluid} with applications to cancer research~\cite{coban2021metastasis, barbazan2023cancer,zubiarrain2024beyond}, but also for more complex
active particles as in pedestrian dynamics~\cite{GuNature2025}, and 
(perhaps much less studied from this side of the research arena), 
for the dynamics of the Earth's mantle~\cite{VoiglanderSM2024}, 
given that both quiescent and active particles constitute the amorphous 
granulated matter on which we step on and build. 
Therefore, we expect our work to be of motivation for many more studies 
in this field and foresee an immediate interest on modeling `mixed' systems of active 
and inactive constituents, subject to external deformations and cyclically changing
conditions as a next step forward.

\section{Methods}
\label{sec:methods}

\subsection{Simulation Details}
We consider a model active glass former composed of a 50:50 bidisperse mixture of $N$ self-propelled soft disks. Particles are located at $\{\dot{\bm r}_i\}_{i=1}^N$ and move in a $L\times L$ box with Lees-Edwards boundary conditions. The dynamics is governed by  the following  overdamped Langevin equations:
\begin{align}
    \begin{split}
        \dot{\bm r}_i &= \mu\bm F_i + \dot{\gamma}y_i\hat{\bm x} + v_0\bm n_i\,,\\
        \dot{\theta}_i &= \sqrt{2D_r}\bm\nu_i\,.\\
    \end{split}
    \label{eq:eom}
\end{align}
The interaction forces between particles $\bm F_i = -\sum_j\nabla V(r_{ij})$ derive from a harmonic pair potential $V(r_{ij}) = \epsilon(1 - d_{ij}/r_{ij})^2\Theta(d_{ij} - r_{ij})$, where $r_{ij}$ is the distance between the centers of particles $i$, $j$ and $d_{ij}$ is the sum of their radii.
Half of the particles have a diameter $d_s=1$ and the others  a diameter $d_b=1.4$. 
Self-propulsion is introduced following the Active Brownian Particles model \cite{fily2012athermal}, namely, including a  constant self-propulsion speed along particle orientations $\bm n_i=(\cos\theta_i,\,\sin\theta_i)$, that diffuses freely with a rotational diffusivity $D_r $.
Such a dynamics introduces persistence in the random motion of the particles, with persistence time $\tau_p=1/D_r$.  We  quantify the strength of activity in the system by a non-dimensional  P\'eclet number Pe$=v_0/(d_sD_r)$, with fixed $\mu=1$. 

The equations of motion~\eqref{eq:eom} are  numerically integrated using a Euler-Maruyama scheme with
a time step $\mathrm dt=0.01$.
The presented results are expressed using the small particle diameter $d_s$ as unit of length, the interaction strength $\epsilon$ as unit of energy and $d_s^2/(\mu\epsilon)$ as unit of time.
Unless otherwise indicated, results are for dense systems at a packing fraction of $\phi=0.9$  (in Fig.~\ref{fig:av-overview}(d) and in the Supplemental Material~\cite{SM} we  provide results on the effect of changing $\phi$).

\subsection{Observables}

A microscopic expression of the stress tensor is provided by the 
 Irving-Kirkwood (IK)  
formula ~\cite{irving1950statistical-mechanical-hydrodynamics}
\begin{equation}
    \sigma_{\alpha\beta}(t) = \frac{1}{2L^2}\sum_{i=1,i\neq j}^N r^\alpha_{ij}(t) F_{ij}^\beta(t)
\end{equation}
where in our $2d$ system Greek indices label cartesian coordinates  $\alpha,\beta\equiv x,\,y$, and Latin ones label particles in the system. 


The participation $P$ is defined as follows. 
Having identified the initial and final times $(t_i,t_f)$ of an avalanche as 
prescribed above, the non-affine displacement of the particles is calculated 
by subtracting the affine term due to simple shear as
\begin{align*}
    \Delta x_i &= x_i(t_f) - x_i(t_i) -\dot{\gamma}(t_f-t_i)y_i(t_f),\\
    \Delta y_i &= y_i(t_f) - y_i(t_i)
\end{align*}
and applying the Lees-Edwards boundary conditions to $\Delta x_i, \Delta y_i$ 
before setting $d_i=\sqrt{\Delta x_i^2 + \Delta y_i^2}$. 
We then measure the so-called ``participation'' $P$ of particles during an 
avalanche $S$,  defined as the inverse fourth moment of the non-affine displacements~\cite{lemaitre2007avalanche-size} 
\begin{equation}
P= \frac{\left(\sum_id_i^2\right)^2}{N\sum_id_i^4}\,.   
\label{eq:participation}
\end{equation}
The number $P$ ranges between $1/N$ (one particle displaces with respect to 
the rest that remain still) and $1$ (identical  displacements of all $N$ 
particles). 

The (fractal) geometry of mobile particles' clusters  is quantified by the radius of gyration
\begin{equation}
    R_g^2=\frac{1}{N_k}\sum_{i\in k}(\boldsymbol{r}_i-\boldsymbol{r}^*_k)^2 \,,
\end{equation}
where $\boldsymbol{r}^*_k$ is the center of mass of a cluster labeled '$k$', made of $N_k$ particles, and the sum runs over all these  particles in the cluster. 

\section*{Acknowledgments}

D.L. and R.W acknowledge MCIU/AEI and DURSI for financial support under Projects No. PID2022-140407NB-C22 and 2021SGR-673, respectively. 
E.E.F. acknowledges support from PIP 2021-2023 CONICET Project Nº~757 and 
the Maria Zambrano program of the Spanish Ministry of Universities through 
the University of Barcelona.







\bibliographystyle{apsrev4-2}
\bibliography{refs}

\begin{thebibliography}{65}%
\makeatletter
\providecommand \@ifxundefined [1]{%
 \@ifx{#1\undefined}
}%
\providecommand \@ifnum [1]{%
 \ifnum #1\expandafter \@firstoftwo
 \else \expandafter \@secondoftwo
 \fi
}%
\providecommand \@ifx [1]{%
 \ifx #1\expandafter \@firstoftwo
 \else \expandafter \@secondoftwo
 \fi
}%
\providecommand \natexlab [1]{#1}%
\providecommand \enquote  [1]{``#1''}%
\providecommand \bibnamefont  [1]{#1}%
\providecommand \bibfnamefont [1]{#1}%
\providecommand \citenamefont [1]{#1}%
\providecommand \href@noop [0]{\@secondoftwo}%
\providecommand \href [0]{\begingroup \@sanitize@url \@href}%
\providecommand \@href[1]{\@@startlink{#1}\@@href}%
\providecommand \@@href[1]{\endgroup#1\@@endlink}%
\providecommand \@sanitize@url [0]{\catcode `\\12\catcode `\$12\catcode
  `\&12\catcode `\#12\catcode `\^12\catcode `\_12\catcode `\%12\relax}%
\providecommand \@@startlink[1]{}%
\providecommand \@@endlink[0]{}%
\providecommand \url  [0]{\begingroup\@sanitize@url \@url }%
\providecommand \@url [1]{\endgroup\@href {#1}{\urlprefix }}%
\providecommand \urlprefix  [0]{URL }%
\providecommand \Eprint [0]{\href }%
\providecommand \doibase [0]{https://doi.org/}%
\providecommand \selectlanguage [0]{\@gobble}%
\providecommand \bibinfo  [0]{\@secondoftwo}%
\providecommand \bibfield  [0]{\@secondoftwo}%
\providecommand \translation [1]{[#1]}%
\providecommand \BibitemOpen [0]{}%
\providecommand \bibitemStop [0]{}%
\providecommand \bibitemNoStop [0]{.\EOS\space}%
\providecommand \EOS [0]{\spacefactor3000\relax}%
\providecommand \BibitemShut  [1]{\csname bibitem#1\endcsname}%
\let\auto@bib@innerbib\@empty
\bibitem [{\citenamefont {Henkes}\ \emph {et~al.}(2011)\citenamefont {Henkes},
  \citenamefont {Fily},\ and\ \citenamefont {Marchetti}}]{henkes2011active}%
  \BibitemOpen
  \bibfield  {author} {\bibinfo {author} {\bibfnamefont {S.}~\bibnamefont
  {Henkes}}, \bibinfo {author} {\bibfnamefont {Y.}~\bibnamefont {Fily}},\ and\
  \bibinfo {author} {\bibfnamefont {M.~C.}\ \bibnamefont {Marchetti}},\ }\href
  {https://doi.org/10.1103/PhysRevE.84.040301} {\bibfield  {journal} {\bibinfo
  {journal} {Phys. Rev. E}\ }\textbf {\bibinfo {volume} {84}},\ \bibinfo
  {pages} {040301} (\bibinfo {year} {2011})}\BibitemShut {NoStop}%
\bibitem [{\citenamefont {Ni}\ \emph {et~al.}(2013)\citenamefont {Ni},
  \citenamefont {Stuart},\ and\ \citenamefont
  {Dijkstra}}]{ni2013pushing-the-glass-transition}%
  \BibitemOpen
  \bibfield  {author} {\bibinfo {author} {\bibfnamefont {R.}~\bibnamefont
  {Ni}}, \bibinfo {author} {\bibfnamefont {M.~A.~C.}\ \bibnamefont {Stuart}},\
  and\ \bibinfo {author} {\bibfnamefont {M.}~\bibnamefont {Dijkstra}},\ }\href
  {https://doi.org/10.1038/ncomms3704} {\bibfield  {journal} {\bibinfo
  {journal} {Nat. Commun.}\ }\textbf {\bibinfo {volume} {4}},\ \bibinfo {pages}
  {2704} (\bibinfo {year} {2013})}\BibitemShut {NoStop}%
\bibitem [{\citenamefont {Berthier}\ and\ \citenamefont
  {Kurchan}(2013)}]{berthier2013non}%
  \BibitemOpen
  \bibfield  {author} {\bibinfo {author} {\bibfnamefont {L.}~\bibnamefont
  {Berthier}}\ and\ \bibinfo {author} {\bibfnamefont {J.}~\bibnamefont
  {Kurchan}},\ }\href {https://doi.org/10.1038/nphys2592} {\bibfield  {journal}
  {\bibinfo  {journal} {Nature Physics}\ }\textbf {\bibinfo {volume} {9}},\
  \bibinfo {pages} {310} (\bibinfo {year} {2013})}\BibitemShut {NoStop}%
\bibitem [{\citenamefont {Berthier}(2014)}]{berthier2014nonequilibrium}%
  \BibitemOpen
  \bibfield  {author} {\bibinfo {author} {\bibfnamefont {L.}~\bibnamefont
  {Berthier}},\ }\href {https://doi.org/10.1103/PhysRevLett.112.220602}
  {\bibfield  {journal} {\bibinfo  {journal} {Phys. Rev. Lett.}\ }\textbf
  {\bibinfo {volume} {112}},\ \bibinfo {pages} {220602} (\bibinfo {year}
  {2014})}\BibitemShut {NoStop}%
\bibitem [{\citenamefont {Mandal}\ \emph {et~al.}(2016)\citenamefont {Mandal},
  \citenamefont {Bhuyan}, \citenamefont {Rao},\ and\ \citenamefont
  {Dasgupta}}]{mandal2016active}%
  \BibitemOpen
  \bibfield  {author} {\bibinfo {author} {\bibfnamefont {R.}~\bibnamefont
  {Mandal}}, \bibinfo {author} {\bibfnamefont {P.~J.}\ \bibnamefont {Bhuyan}},
  \bibinfo {author} {\bibfnamefont {M.}~\bibnamefont {Rao}},\ and\ \bibinfo
  {author} {\bibfnamefont {C.}~\bibnamefont {Dasgupta}},\ }\href
  {https://doi.org/10.1039/C5SM02950C} {\bibfield  {journal} {\bibinfo
  {journal} {Soft Matter}\ }\textbf {\bibinfo {volume} {12}},\ \bibinfo {pages}
  {6268} (\bibinfo {year} {2016})}\BibitemShut {NoStop}%
\bibitem [{\citenamefont {Ding}\ \emph {et~al.}(2017)\citenamefont {Ding},
  \citenamefont {Jiang},\ and\ \citenamefont {Hou}}]{ding2017study}%
  \BibitemOpen
  \bibfield  {author} {\bibinfo {author} {\bibfnamefont {H.}~\bibnamefont
  {Ding}}, \bibinfo {author} {\bibfnamefont {H.}~\bibnamefont {Jiang}},\ and\
  \bibinfo {author} {\bibfnamefont {Z.}~\bibnamefont {Hou}},\ }\href
  {https://doi.org/10.1103/physreve.95.052608} {\bibfield  {journal} {\bibinfo
  {journal} {Phys. Rev. E}\ }\textbf {\bibinfo {volume} {95}},\ \bibinfo
  {pages} {052608} (\bibinfo {year} {2017})}\BibitemShut {NoStop}%
\bibitem [{\citenamefont {Liluashvili}\ \emph {et~al.}(2017)\citenamefont
  {Liluashvili}, \citenamefont {\'Onody},\ and\ \citenamefont
  {Voigtmann}}]{liluashvili2017mode}%
  \BibitemOpen
  \bibfield  {author} {\bibinfo {author} {\bibfnamefont {A.}~\bibnamefont
  {Liluashvili}}, \bibinfo {author} {\bibfnamefont {J.}~\bibnamefont
  {\'Onody}},\ and\ \bibinfo {author} {\bibfnamefont {T.}~\bibnamefont
  {Voigtmann}},\ }\href {https://doi.org/10.1103/PhysRevE.96.062608} {\bibfield
   {journal} {\bibinfo  {journal} {Phys. Rev. E}\ }\textbf {\bibinfo {volume}
  {96}},\ \bibinfo {pages} {062608} (\bibinfo {year} {2017})}\BibitemShut
  {NoStop}%
\bibitem [{\citenamefont {Janssen}(2019)}]{janssen2019active}%
  \BibitemOpen
  \bibfield  {author} {\bibinfo {author} {\bibfnamefont {L.~M.~C.}\
  \bibnamefont {Janssen}},\ }\href {https://doi.org/10.1088/1361-648X/ab3e90}
  {\bibfield  {journal} {\bibinfo  {journal} {Journal of Physics: Condensed
  Matter}\ }\textbf {\bibinfo {volume} {31}},\ \bibinfo {pages} {503002}
  (\bibinfo {year} {2019})}\BibitemShut {NoStop}%
\bibitem [{\citenamefont {Berthier}\ \emph {et~al.}(2019)\citenamefont
  {Berthier}, \citenamefont {Flenner},\ and\ \citenamefont
  {Szamel}}]{berthier2019glassy}%
  \BibitemOpen
  \bibfield  {author} {\bibinfo {author} {\bibfnamefont {L.}~\bibnamefont
  {Berthier}}, \bibinfo {author} {\bibfnamefont {E.}~\bibnamefont {Flenner}},\
  and\ \bibinfo {author} {\bibfnamefont {G.}~\bibnamefont {Szamel}},\ }\href
  {https://doi.org/10.1063/1.5093240} {\bibfield  {journal} {\bibinfo
  {journal} {The Journal of Chemical Physics}\ }\textbf {\bibinfo {volume}
  {150}},\ \bibinfo {pages} {200901} (\bibinfo {year} {2019})}\BibitemShut
  {NoStop}%
\bibitem [{\citenamefont {Henkes}\ \emph {et~al.}(2020)\citenamefont {Henkes},
  \citenamefont {Kostanjevec}, \citenamefont {Collinson}, \citenamefont
  {Sknepnek},\ and\ \citenamefont {Bertin}}]{henkes2020dense}%
  \BibitemOpen
  \bibfield  {author} {\bibinfo {author} {\bibfnamefont {S.}~\bibnamefont
  {Henkes}}, \bibinfo {author} {\bibfnamefont {K.}~\bibnamefont {Kostanjevec}},
  \bibinfo {author} {\bibfnamefont {J.~M.}\ \bibnamefont {Collinson}}, \bibinfo
  {author} {\bibfnamefont {R.}~\bibnamefont {Sknepnek}},\ and\ \bibinfo
  {author} {\bibfnamefont {E.}~\bibnamefont {Bertin}},\ }\href
  {https://doi.org/10.1038/s41467-020-15164-5} {\bibfield  {journal} {\bibinfo
  {journal} {Nature Communications}\ }\textbf {\bibinfo {volume} {11}},\
  \bibinfo {pages} {1405} (\bibinfo {year} {2020})}\BibitemShut {NoStop}%
\bibitem [{\citenamefont {Paoluzzi}\ \emph {et~al.}(2022)\citenamefont
  {Paoluzzi}, \citenamefont {Levis},\ and\ \citenamefont
  {Pagonabarraga}}]{paoluzzi2022motility}%
  \BibitemOpen
  \bibfield  {author} {\bibinfo {author} {\bibfnamefont {M.}~\bibnamefont
  {Paoluzzi}}, \bibinfo {author} {\bibfnamefont {D.}~\bibnamefont {Levis}},\
  and\ \bibinfo {author} {\bibfnamefont {I.}~\bibnamefont {Pagonabarraga}},\
  }\href {https://doi.org/10.1038/s42005-022-00886-3} {\bibfield  {journal}
  {\bibinfo  {journal} {Communications Physics}\ }\textbf {\bibinfo {volume}
  {5}},\ \bibinfo {pages} {111} (\bibinfo {year} {2022})}\BibitemShut {NoStop}%
\bibitem [{\citenamefont {Keta}\ \emph {et~al.}(2022)\citenamefont {Keta},
  \citenamefont {Jack},\ and\ \citenamefont {Berthier}}]{keta2022disordered}%
  \BibitemOpen
  \bibfield  {author} {\bibinfo {author} {\bibfnamefont {Y.-E.}\ \bibnamefont
  {Keta}}, \bibinfo {author} {\bibfnamefont {R.~L.}\ \bibnamefont {Jack}},\
  and\ \bibinfo {author} {\bibfnamefont {L.}~\bibnamefont {Berthier}},\ }\href
  {https://doi.org/10.1103/PhysRevLett.129.048002} {\bibfield  {journal}
  {\bibinfo  {journal} {Phys. Rev. Lett.}\ }\textbf {\bibinfo {volume} {129}},\
  \bibinfo {pages} {048002} (\bibinfo {year} {2022})}\BibitemShut {NoStop}%
\bibitem [{\citenamefont {Sadhukhan}\ \emph {et~al.}(2024)\citenamefont
  {Sadhukhan}, \citenamefont {Dey}, \citenamefont {Karmakar},\ and\
  \citenamefont {Nandi}}]{karmakar2024perspective-active-glassy}%
  \BibitemOpen
  \bibfield  {author} {\bibinfo {author} {\bibfnamefont {S.}~\bibnamefont
  {Sadhukhan}}, \bibinfo {author} {\bibfnamefont {S.}~\bibnamefont {Dey}},
  \bibinfo {author} {\bibfnamefont {S.}~\bibnamefont {Karmakar}},\ and\
  \bibinfo {author} {\bibfnamefont {S.~K.}\ \bibnamefont {Nandi}},\ }\href
  {http://dx.doi.org/10.1140/epjs/s11734-024-01188-1} {\bibfield  {journal}
  {\bibinfo  {journal} {Euro. Phys. J. Special Topics}\ } (\bibinfo {year}
  {2024})}\BibitemShut {NoStop}%
\bibitem [{\citenamefont {Ghosh}\ \emph {et~al.}(2025)\citenamefont {Ghosh},
  \citenamefont {Sollich},\ and\ \citenamefont
  {Nandi}}]{sollich2025elastoplastic-model}%
  \BibitemOpen
  \bibfield  {author} {\bibinfo {author} {\bibfnamefont {T.}~\bibnamefont
  {Ghosh}}, \bibinfo {author} {\bibfnamefont {P.}~\bibnamefont {Sollich}},\
  and\ \bibinfo {author} {\bibfnamefont {S.~K.}\ \bibnamefont {Nandi}},\
  }\bibfield  {journal} {\bibinfo  {journal} {Soft Matter}\ }\href
  {https://doi.org/10.1039/d4sm01394h} {10.1039/d4sm01394h} (\bibinfo {year}
  {2025})\BibitemShut {NoStop}%
\bibitem [{\citenamefont {Goswami}\ \emph {et~al.}(2025)\citenamefont
  {Goswami}, \citenamefont {Shivashankar},\ and\ \citenamefont
  {Sastry}}]{GoswamiNP2025}%
  \BibitemOpen
  \bibfield  {author} {\bibinfo {author} {\bibfnamefont {Y.}~\bibnamefont
  {Goswami}}, \bibinfo {author} {\bibfnamefont {G.~V.}\ \bibnamefont
  {Shivashankar}},\ and\ \bibinfo {author} {\bibfnamefont {S.}~\bibnamefont
  {Sastry}},\ }\href {https://doi.org/10.1038/s41567-025-02843-7} {\bibfield
  {journal} {\bibinfo  {journal} {Nature Physics}\ }\textbf {\bibinfo {volume}
  {21}},\ \bibinfo {pages} {817} (\bibinfo {year} {2025})}\BibitemShut
  {NoStop}%
\bibitem [{\citenamefont {Maloney}\ and\ \citenamefont
  {Lema{\^i}tre}(2004)}]{maloney2004subextensive}%
  \BibitemOpen
  \bibfield  {author} {\bibinfo {author} {\bibfnamefont {C.}~\bibnamefont
  {Maloney}}\ and\ \bibinfo {author} {\bibfnamefont {A.}~\bibnamefont
  {Lema{\^i}tre}},\ }\href {http://dx.doi.org/10.1103/physrevlett.93.016001}
  {\bibfield  {journal} {\bibinfo  {journal} {Phys. Rev. Lett.}\ }\textbf
  {\bibinfo {volume} {93}},\ \bibinfo {pages} {016001} (\bibinfo {year}
  {2004})}\BibitemShut {NoStop}%
\bibitem [{\citenamefont {Maloney}\ and\ \citenamefont
  {Lema{\^i}tre}(2006)}]{maloney2006amorphous}%
  \BibitemOpen
  \bibfield  {author} {\bibinfo {author} {\bibfnamefont {C.~E.}\ \bibnamefont
  {Maloney}}\ and\ \bibinfo {author} {\bibfnamefont {A.}~\bibnamefont
  {Lema{\^i}tre}},\ }\href {http://dx.doi.org/10.1103/physreve.74.016118}
  {\bibfield  {journal} {\bibinfo  {journal} {Phys. Rev. E}\ }\textbf {\bibinfo
  {volume} {74}},\ \bibinfo {pages} {016118} (\bibinfo {year}
  {2006})}\BibitemShut {NoStop}%
\bibitem [{\citenamefont {Salerno}\ \emph {et~al.}(2012)\citenamefont
  {Salerno}, \citenamefont {Maloney},\ and\ \citenamefont
  {Robbins}}]{robbins2012avalanches}%
  \BibitemOpen
  \bibfield  {author} {\bibinfo {author} {\bibfnamefont {K.~M.}\ \bibnamefont
  {Salerno}}, \bibinfo {author} {\bibfnamefont {C.~E.}\ \bibnamefont
  {Maloney}},\ and\ \bibinfo {author} {\bibfnamefont {M.~O.}\ \bibnamefont
  {Robbins}},\ }\href {https://doi.org/10.1103/physrevlett.109.105703}
  {\bibfield  {journal} {\bibinfo  {journal} {Phys. Rev. Lett.}\ }\textbf
  {\bibinfo {volume} {109}},\ \bibinfo {pages} {105703} (\bibinfo {year}
  {2012})}\BibitemShut {NoStop}%
\bibitem [{\citenamefont {Salerno}\ and\ \citenamefont
  {Robbins}(2013)}]{salerno2013effect-of-inertia}%
  \BibitemOpen
  \bibfield  {author} {\bibinfo {author} {\bibfnamefont {K.~M.}\ \bibnamefont
  {Salerno}}\ and\ \bibinfo {author} {\bibfnamefont {M.~O.}\ \bibnamefont
  {Robbins}},\ }\href {https://doi.org/10.1103/physreve.88.062206} {\bibfield
  {journal} {\bibinfo  {journal} {Phys. Rev. E}\ }\textbf {\bibinfo {volume}
  {88}},\ \bibinfo {pages} {062206} (\bibinfo {year} {2013})}\BibitemShut
  {NoStop}%
\bibitem [{\citenamefont {Liu}\ \emph {et~al.}(2016)\citenamefont {Liu},
  \citenamefont {Ferrero}, \citenamefont {Puosi}, \citenamefont {Barrat},\ and\
  \citenamefont {Martens}}]{LiuPRL2016}%
  \BibitemOpen
  \bibfield  {author} {\bibinfo {author} {\bibfnamefont {C.}~\bibnamefont
  {Liu}}, \bibinfo {author} {\bibfnamefont {E.~E.}\ \bibnamefont {Ferrero}},
  \bibinfo {author} {\bibfnamefont {F.}~\bibnamefont {Puosi}}, \bibinfo
  {author} {\bibfnamefont {J.-L.}\ \bibnamefont {Barrat}},\ and\ \bibinfo
  {author} {\bibfnamefont {K.}~\bibnamefont {Martens}},\ }\href
  {https://doi.org/10.1103/PhysRevLett.116.065501} {\bibfield  {journal}
  {\bibinfo  {journal} {Phys. Rev. Lett.}\ }\textbf {\bibinfo {volume} {116}},\
  \bibinfo {pages} {065501} (\bibinfo {year} {2016})}\BibitemShut {NoStop}%
\bibitem [{\citenamefont {Zhang}\ \emph {et~al.}(2017)\citenamefont {Zhang},
  \citenamefont {Dahmen},\ and\ \citenamefont
  {Ostoja-Starzewski}}]{dahmen2017scaling}%
  \BibitemOpen
  \bibfield  {author} {\bibinfo {author} {\bibfnamefont {D.}~\bibnamefont
  {Zhang}}, \bibinfo {author} {\bibfnamefont {K.~A.}\ \bibnamefont {Dahmen}},\
  and\ \bibinfo {author} {\bibfnamefont {M.}~\bibnamefont
  {Ostoja-Starzewski}},\ }\href {https://doi.org/10.1103/physreve.95.032902}
  {\bibfield  {journal} {\bibinfo  {journal} {Phys. Rev. E}\ }\textbf {\bibinfo
  {volume} {95}},\ \bibinfo {pages} {032902} (\bibinfo {year}
  {2017})}\BibitemShut {NoStop}%
\bibitem [{\citenamefont {Ruscher}\ and\ \citenamefont
  {Rottler}(2021)}]{RuscherTL2021}%
  \BibitemOpen
  \bibfield  {author} {\bibinfo {author} {\bibfnamefont {C.}~\bibnamefont
  {Ruscher}}\ and\ \bibinfo {author} {\bibfnamefont {J.}~\bibnamefont
  {Rottler}},\ }\href {https://doi.org/10.1007/s11249-021-01439-5} {\bibfield
  {journal} {\bibinfo  {journal} {Tribology Letters}\ }\textbf {\bibinfo
  {volume} {69}},\ \bibinfo {pages} {64} (\bibinfo {year} {2021})}\BibitemShut
  {NoStop}%
\bibitem [{\citenamefont {Clemmer}\ \emph {et~al.}(2021)\citenamefont
  {Clemmer}, \citenamefont {Salerno},\ and\ \citenamefont
  {Robbins}}]{clemmer2021criticality-II}%
  \BibitemOpen
  \bibfield  {author} {\bibinfo {author} {\bibfnamefont {J.~T.}\ \bibnamefont
  {Clemmer}}, \bibinfo {author} {\bibfnamefont {K.~M.}\ \bibnamefont
  {Salerno}},\ and\ \bibinfo {author} {\bibfnamefont {M.~O.}\ \bibnamefont
  {Robbins}},\ }\href {https://doi.org/10.1103/physreve.103.042606} {\bibfield
  {journal} {\bibinfo  {journal} {Phys. Rev. E}\ }\textbf {\bibinfo {volume}
  {103}},\ \bibinfo {pages} {042606} (\bibinfo {year} {2021})}\BibitemShut
  {NoStop}%
\bibitem [{\citenamefont {Saitoh}(2025)}]{SaitohFP2025}%
  \BibitemOpen
  \bibfield  {author} {\bibinfo {author} {\bibfnamefont {K.}~\bibnamefont
  {Saitoh}},\ }\href {https://doi.org/10.3389/fphy.2025.1548966} {\bibfield
  {journal} {\bibinfo  {journal} {Frontiers in Physics}\ }\textbf {\bibinfo
  {volume} {13}},\ \bibinfo {pages} {1548966} (\bibinfo {year}
  {2025})}\BibitemShut {NoStop}%
\bibitem [{\citenamefont {Oyama}\ \emph {et~al.}(2021)\citenamefont {Oyama},
  \citenamefont {Mizuno},\ and\ \citenamefont {Ikeda}}]{OyamaPRE2021}%
  \BibitemOpen
  \bibfield  {author} {\bibinfo {author} {\bibfnamefont {N.}~\bibnamefont
  {Oyama}}, \bibinfo {author} {\bibfnamefont {H.}~\bibnamefont {Mizuno}},\ and\
  \bibinfo {author} {\bibfnamefont {A.}~\bibnamefont {Ikeda}},\ }\href
  {https://doi.org/10.1103/PhysRevE.104.015002} {\bibfield  {journal} {\bibinfo
   {journal} {Phys. Rev. E}\ }\textbf {\bibinfo {volume} {104}},\ \bibinfo
  {pages} {015002} (\bibinfo {year} {2021})}\BibitemShut {NoStop}%
\bibitem [{\citenamefont {Morse}\ \emph {et~al.}(2021)\citenamefont {Morse},
  \citenamefont {Roy}, \citenamefont {Agoritsas}, \citenamefont {Stanifer},
  \citenamefont {Corwin},\ and\ \citenamefont
  {Manning}}]{morse2021link-active-matter-sheared-granular}%
  \BibitemOpen
  \bibfield  {author} {\bibinfo {author} {\bibfnamefont {P.~K.}\ \bibnamefont
  {Morse}}, \bibinfo {author} {\bibfnamefont {S.}~\bibnamefont {Roy}}, \bibinfo
  {author} {\bibfnamefont {E.}~\bibnamefont {Agoritsas}}, \bibinfo {author}
  {\bibfnamefont {E.}~\bibnamefont {Stanifer}}, \bibinfo {author}
  {\bibfnamefont {E.~I.}\ \bibnamefont {Corwin}},\ and\ \bibinfo {author}
  {\bibfnamefont {M.~L.}\ \bibnamefont {Manning}},\ }\href
  {http://dx.doi.org/10.1073/pnas.2019909118} {\bibfield  {journal} {\bibinfo
  {journal} {Proc. Natl. Acad. Sci.}\ }\textbf {\bibinfo {volume} {118}},\
  \bibinfo {pages} {e2019909118} (\bibinfo {year} {2021})}\BibitemShut
  {NoStop}%
\bibitem [{\citenamefont {Keta}\ \emph {et~al.}(2023)\citenamefont {Keta},
  \citenamefont {Mandal}, \citenamefont {Sollich}, \citenamefont {Jack},\ and\
  \citenamefont {Berthier}}]{keta2023intermittent-relaxation}%
  \BibitemOpen
  \bibfield  {author} {\bibinfo {author} {\bibfnamefont {Y.-E.}\ \bibnamefont
  {Keta}}, \bibinfo {author} {\bibfnamefont {R.}~\bibnamefont {Mandal}},
  \bibinfo {author} {\bibfnamefont {P.}~\bibnamefont {Sollich}}, \bibinfo
  {author} {\bibfnamefont {R.~L.}\ \bibnamefont {Jack}},\ and\ \bibinfo
  {author} {\bibfnamefont {L.}~\bibnamefont {Berthier}},\ }\href
  {https://doi.org/10.1039/d3sm00034f} {\bibfield  {journal} {\bibinfo
  {journal} {Soft Matter}\ }\textbf {\bibinfo {volume} {19}},\ \bibinfo {pages}
  {3871–3883} (\bibinfo {year} {2023})}\BibitemShut {NoStop}%
\bibitem [{\citenamefont {Villarroel}\ and\ \citenamefont
  {D{\"u}ring}(2024)}]{villarroel2024avalanche-properties}%
  \BibitemOpen
  \bibfield  {author} {\bibinfo {author} {\bibfnamefont {C.}~\bibnamefont
  {Villarroel}}\ and\ \bibinfo {author} {\bibfnamefont {G.}~\bibnamefont
  {D{\"u}ring}},\ }\href {https://doi.org/10.1039/d3sm01354e} {\bibfield
  {journal} {\bibinfo  {journal} {Soft Matter}\ }\textbf {\bibinfo {volume}
  {20}},\ \bibinfo {pages} {3520–3528} (\bibinfo {year} {2024})}\BibitemShut
  {NoStop}%
\bibitem [{\citenamefont {Reichhardt}\ and\ \citenamefont
  {Reichhardt}(2018)}]{ReichhardtNJP2018}%
  \BibitemOpen
  \bibfield  {author} {\bibinfo {author} {\bibfnamefont {C.~J.~O.}\
  \bibnamefont {Reichhardt}}\ and\ \bibinfo {author} {\bibfnamefont
  {C.}~\bibnamefont {Reichhardt}},\ }\href
  {https://doi.org/10.1088/1367-2630/aaa392} {\bibfield  {journal} {\bibinfo
  {journal} {New Journal of Physics}\ }\textbf {\bibinfo {volume} {20}},\
  \bibinfo {pages} {025002} (\bibinfo {year} {2018})}\BibitemShut {NoStop}%
\bibitem [{\citenamefont {Villarroel}\ and\ \citenamefont
  {D{\"u}ring}(2021)}]{villarroel2021critical-yielding-rheology}%
  \BibitemOpen
  \bibfield  {author} {\bibinfo {author} {\bibfnamefont {C.}~\bibnamefont
  {Villarroel}}\ and\ \bibinfo {author} {\bibfnamefont {G.}~\bibnamefont
  {D{\"u}ring}},\ }\href {http://dx.doi.org/10.1039/d1sm00948f} {\bibfield
  {journal} {\bibinfo  {journal} {Soft Matter}\ }\textbf {\bibinfo {volume}
  {17}},\ \bibinfo {pages} {9944–9949} (\bibinfo {year} {2021})}\BibitemShut
  {NoStop}%
\bibitem [{\citenamefont {Wiese}\ \emph {et~al.}(2023)\citenamefont {Wiese},
  \citenamefont {Kroy},\ and\ \citenamefont {Levis}}]{WiesePRL2023}%
  \BibitemOpen
  \bibfield  {author} {\bibinfo {author} {\bibfnamefont {R.}~\bibnamefont
  {Wiese}}, \bibinfo {author} {\bibfnamefont {K.}~\bibnamefont {Kroy}},\ and\
  \bibinfo {author} {\bibfnamefont {D.}~\bibnamefont {Levis}},\ }\href
  {https://doi.org/10.1103/PhysRevLett.131.178302} {\bibfield  {journal}
  {\bibinfo  {journal} {Phys. Rev. Lett.}\ }\textbf {\bibinfo {volume} {131}},\
  \bibinfo {pages} {178302} (\bibinfo {year} {2023})}\BibitemShut {NoStop}%
\bibitem [{\citenamefont {Cates}\ and\ \citenamefont
  {Tailleur}(2015)}]{cates2015motility}%
  \BibitemOpen
  \bibfield  {author} {\bibinfo {author} {\bibfnamefont {M.~E.}\ \bibnamefont
  {Cates}}\ and\ \bibinfo {author} {\bibfnamefont {J.}~\bibnamefont
  {Tailleur}},\ }\href
  {https://doi.org/https://doi.org/10.1146/annurev-conmatphys-031214-014710}
  {\bibfield  {journal} {\bibinfo  {journal} {Annual Review of Condensed Matter
  Physics}\ }\textbf {\bibinfo {volume} {6}},\ \bibinfo {pages} {219} (\bibinfo
  {year} {2015})}\BibitemShut {NoStop}%
\bibitem [{\citenamefont {Bialk{\'e}}\ \emph {et~al.}(2013)\citenamefont
  {Bialk{\'e}}, \citenamefont {Löwen},\ and\ \citenamefont
  {Speck}}]{bialke2013microscopic}%
  \BibitemOpen
  \bibfield  {author} {\bibinfo {author} {\bibfnamefont {J.}~\bibnamefont
  {Bialk{\'e}}}, \bibinfo {author} {\bibfnamefont {H.}~\bibnamefont {Löwen}},\
  and\ \bibinfo {author} {\bibfnamefont {T.}~\bibnamefont {Speck}},\ }\href
  {https://doi.org/10.1209/0295-5075/103/30008} {\bibfield  {journal} {\bibinfo
   {journal} {Europhysics Letters}\ }\textbf {\bibinfo {volume} {103}},\
  \bibinfo {pages} {30008} (\bibinfo {year} {2013})}\BibitemShut {NoStop}%
\bibitem [{\citenamefont {Farage}\ \emph {et~al.}(2015)\citenamefont {Farage},
  \citenamefont {Krinninger},\ and\ \citenamefont
  {Brader}}]{farage2015effective}%
  \BibitemOpen
  \bibfield  {author} {\bibinfo {author} {\bibfnamefont {T.~F.~F.}\
  \bibnamefont {Farage}}, \bibinfo {author} {\bibfnamefont {P.}~\bibnamefont
  {Krinninger}},\ and\ \bibinfo {author} {\bibfnamefont {J.~M.}\ \bibnamefont
  {Brader}},\ }\href {https://doi.org/10.1103/PhysRevE.91.042310} {\bibfield
  {journal} {\bibinfo  {journal} {Phys. Rev. E}\ }\textbf {\bibinfo {volume}
  {91}},\ \bibinfo {pages} {042310} (\bibinfo {year} {2015})}\BibitemShut
  {NoStop}%
\bibitem [{\citenamefont {Ses\'e-Sansa}\ \emph {et~al.}(2021)\citenamefont
  {Ses\'e-Sansa}, \citenamefont {Levis},\ and\ \citenamefont
  {Pagonabarraga}}]{sese2021phase}%
  \BibitemOpen
  \bibfield  {author} {\bibinfo {author} {\bibfnamefont {E.}~\bibnamefont
  {Ses\'e-Sansa}}, \bibinfo {author} {\bibfnamefont {D.}~\bibnamefont
  {Levis}},\ and\ \bibinfo {author} {\bibfnamefont {I.}~\bibnamefont
  {Pagonabarraga}},\ }\href {https://doi.org/10.1103/PhysRevE.104.054611}
  {\bibfield  {journal} {\bibinfo  {journal} {Phys. Rev. E}\ }\textbf {\bibinfo
  {volume} {104}},\ \bibinfo {pages} {054611} (\bibinfo {year}
  {2021})}\BibitemShut {NoStop}%
\bibitem [{\citenamefont {Agoritsas}(2021)}]{agoritsas2021mean-field}%
  \BibitemOpen
  \bibfield  {author} {\bibinfo {author} {\bibfnamefont {E.}~\bibnamefont
  {Agoritsas}},\ }\href {https://doi.org/10.1088/1742-5468/abdd18} {\bibfield
  {journal} {\bibinfo  {journal} {J. Stat. Mech.}\ }\textbf {\bibinfo {volume}
  {2021}},\ \bibinfo {pages} {033501} (\bibinfo {year} {2021})}\BibitemShut
  {NoStop}%
\bibitem [{\citenamefont {Lees}\ and\ \citenamefont
  {Edwards}(1972)}]{lees1972computer-study-of-transport-processes}%
  \BibitemOpen
  \bibfield  {author} {\bibinfo {author} {\bibfnamefont {A.~W.}\ \bibnamefont
  {Lees}}\ and\ \bibinfo {author} {\bibfnamefont {S.~F.}\ \bibnamefont
  {Edwards}},\ }\href {https://doi.org/10.1088/0022-3719/5/15/006} {\bibfield
  {journal} {\bibinfo  {journal} {J. Phys. C: Solid State Phys.}\ }\textbf
  {\bibinfo {volume} {5}},\ \bibinfo {pages} {1921–1928} (\bibinfo {year}
  {1972})}\BibitemShut {NoStop}%
\bibitem [{\citenamefont {Allen}\ and\ \citenamefont
  {Tildesley}(2017)}]{allen-tildesley2017computer-simulation-of-liquids}%
  \BibitemOpen
  \bibfield  {author} {\bibinfo {author} {\bibfnamefont {M.~P.}\ \bibnamefont
  {Allen}}\ and\ \bibinfo {author} {\bibfnamefont {D.~J.}\ \bibnamefont
  {Tildesley}},\ }\href {https://doi.org/10.1093/oso/9780198803195.001.0001}
  {\emph {\bibinfo {title} {Computer {S}imulation of {L}iquids}}}\ (\bibinfo
  {publisher} {Oxford University Press},\ \bibinfo {year} {2017})\BibitemShut
  {NoStop}%
\bibitem [{\citenamefont {Nicolas}\ \emph {et~al.}(2018)\citenamefont
  {Nicolas}, \citenamefont {Ferrero}, \citenamefont {Martens},\ and\
  \citenamefont {Barrat}}]{NicolasRMP2018}%
  \BibitemOpen
  \bibfield  {author} {\bibinfo {author} {\bibfnamefont {A.}~\bibnamefont
  {Nicolas}}, \bibinfo {author} {\bibfnamefont {E.~E.}\ \bibnamefont
  {Ferrero}}, \bibinfo {author} {\bibfnamefont {K.}~\bibnamefont {Martens}},\
  and\ \bibinfo {author} {\bibfnamefont {J.-L.}\ \bibnamefont {Barrat}},\
  }\href {https://doi.org/10.1103/RevModPhys.90.045006} {\bibfield  {journal}
  {\bibinfo  {journal} {Rev. Mod. Phys.}\ }\textbf {\bibinfo {volume} {90}},\
  \bibinfo {pages} {045006} (\bibinfo {year} {2018})}\BibitemShut {NoStop}%
\bibitem [{\citenamefont {Navas-Portella}\ \emph {et~al.}(2016)\citenamefont
  {Navas-Portella}, \citenamefont {Corral},\ and\ \citenamefont
  {Vives}}]{VivesPRE2016}%
  \BibitemOpen
  \bibfield  {author} {\bibinfo {author} {\bibfnamefont {V.}~\bibnamefont
  {Navas-Portella}}, \bibinfo {author} {\bibfnamefont {A.}~\bibnamefont
  {Corral}},\ and\ \bibinfo {author} {\bibfnamefont {E.}~\bibnamefont
  {Vives}},\ }\href {https://doi.org/10.1103/PhysRevE.94.033005} {\bibfield
  {journal} {\bibinfo  {journal} {Phys. Rev. E}\ }\textbf {\bibinfo {volume}
  {94}},\ \bibinfo {pages} {033005} (\bibinfo {year} {2016})}\BibitemShut
  {NoStop}%
\bibitem [{\citenamefont {Jani\ifmmode \acute{c}\else
  \'{c}\fi{}evi\ifmmode~\acute{c}\else \'{c}\fi{}}\ \emph
  {et~al.}(2016)\citenamefont {Jani\ifmmode \acute{c}\else
  \'{c}\fi{}evi\ifmmode~\acute{c}\else \'{c}\fi{}}, \citenamefont {Laurson},
  \citenamefont {M\aa{}l\o{}y}, \citenamefont {Santucci},\ and\ \citenamefont
  {Alava}}]{AlavaPRL2016}%
  \BibitemOpen
  \bibfield  {author} {\bibinfo {author} {\bibfnamefont {S.}~\bibnamefont
  {Jani\ifmmode \acute{c}\else \'{c}\fi{}evi\ifmmode~\acute{c}\else
  \'{c}\fi{}}}, \bibinfo {author} {\bibfnamefont {L.}~\bibnamefont {Laurson}},
  \bibinfo {author} {\bibfnamefont {K.~J.}\ \bibnamefont {M\aa{}l\o{}y}},
  \bibinfo {author} {\bibfnamefont {S.}~\bibnamefont {Santucci}},\ and\
  \bibinfo {author} {\bibfnamefont {M.~J.}\ \bibnamefont {Alava}},\ }\href
  {https://doi.org/10.1103/PhysRevLett.117.230601} {\bibfield  {journal}
  {\bibinfo  {journal} {Phys. Rev. Lett.}\ }\textbf {\bibinfo {volume} {117}},\
  \bibinfo {pages} {230601} (\bibinfo {year} {2016})}\BibitemShut {NoStop}%
\bibitem [{\citenamefont {Villegas}\ \emph {et~al.}(2019)\citenamefont
  {Villegas}, \citenamefont {di~Santo}, \citenamefont {Burioni},\ and\
  \citenamefont {Mu\~noz}}]{MunozPRE2019}%
  \BibitemOpen
  \bibfield  {author} {\bibinfo {author} {\bibfnamefont {P.}~\bibnamefont
  {Villegas}}, \bibinfo {author} {\bibfnamefont {S.}~\bibnamefont {di~Santo}},
  \bibinfo {author} {\bibfnamefont {R.}~\bibnamefont {Burioni}},\ and\ \bibinfo
  {author} {\bibfnamefont {M.~A.}\ \bibnamefont {Mu\~noz}},\ }\href
  {https://doi.org/10.1103/PhysRevE.100.012133} {\bibfield  {journal} {\bibinfo
   {journal} {Phys. Rev. E}\ }\textbf {\bibinfo {volume} {100}},\ \bibinfo
  {pages} {012133} (\bibinfo {year} {2019})}\BibitemShut {NoStop}%
\bibitem [{Note1()}]{Note1}%
  \BibitemOpen
  \bibinfo {note} {We note that both the time step $\Delta t$ and the threshold
  used to define avalanches affect numerical details of the distributions.
  Variations of the latter mainly shift the power-law regimes discussed in the
  next section, without altering the scaling exponent. For consistency, we use
  a zero threshold, to avoid parameter optimization, while keeping the results
  stable.}\BibitemShut {Stop}%
\bibitem [{SM(2025)}]{SM}%
  \BibitemOpen
  \href@noop {} {\bibinfo {title} {See supplemental material at \emph{URL will
  be inserted by publisher} for more details.}} (\bibinfo {year}
  {2025})\BibitemShut {NoStop}%
\bibitem [{Note2()}]{Note2}%
  \BibitemOpen
  \bibinfo {note} {The critical aspect of the yielding transition is better
  seen when considering the strain rate as an order parameter and the stress as
  the control parameter~\cite {FerreroSM2019, NicolasRMP2018}}\BibitemShut
  {NoStop}%
\bibitem [{Note3()}]{Note3}%
  \BibitemOpen
  \bibinfo {note} {Even though we have tried to make the analysis of both
  stress signals compatible, e.g., by detecting the avalanche time windows in
  $\sigma (t)$ and using the same starting/ending points for both $\sigma (t)$
  and $\sigma _R(t)$.}\BibitemShut {Stop}%
\bibitem [{Note4()}]{Note4}%
  \BibitemOpen
  \bibinfo {note} {To obtain the results shown in Fig.~\ref {fig:av-random} we
  have changed the persistence time $\tau _p$ while keeping a fixed velocity of
  self-propulsion $v_0=0.009$, which amounts to $\protect \mathrm {Pe}=30$ for
  $\tau _p=3333$. This was done to avoid the active force becoming stronger
  than the steric repulsion forces. Otherwise, artifacts could arise in the
  thermal limit of small persistence when the two forces become comparable,
  allowing particles to penetrate each other.}\BibitemShut {Stop}%
\bibitem [{\citenamefont {Talamali}\ \emph {et~al.}(2011)\citenamefont
  {Talamali}, \citenamefont {Pet\"aj\"a}, \citenamefont {Vandembroucq},\ and\
  \citenamefont {Roux}}]{Talamali2011}%
  \BibitemOpen
  \bibfield  {author} {\bibinfo {author} {\bibfnamefont {M.}~\bibnamefont
  {Talamali}}, \bibinfo {author} {\bibfnamefont {V.}~\bibnamefont
  {Pet\"aj\"a}}, \bibinfo {author} {\bibfnamefont {D.}~\bibnamefont
  {Vandembroucq}},\ and\ \bibinfo {author} {\bibfnamefont {S.}~\bibnamefont
  {Roux}},\ }\href {https://doi.org/10.1103/PhysRevE.84.016115} {\bibfield
  {journal} {\bibinfo  {journal} {Phys. Rev. E}\ }\textbf {\bibinfo {volume}
  {84}},\ \bibinfo {pages} {016115} (\bibinfo {year} {2011})}\BibitemShut
  {NoStop}%
\bibitem [{\citenamefont {Lin}\ \emph {et~al.}(2014)\citenamefont {Lin},
  \citenamefont {Lerner}, \citenamefont {Rosso},\ and\ \citenamefont
  {Wyart}}]{LinPNAS2014}%
  \BibitemOpen
  \bibfield  {author} {\bibinfo {author} {\bibfnamefont {J.}~\bibnamefont
  {Lin}}, \bibinfo {author} {\bibfnamefont {E.}~\bibnamefont {Lerner}},
  \bibinfo {author} {\bibfnamefont {A.}~\bibnamefont {Rosso}},\ and\ \bibinfo
  {author} {\bibfnamefont {M.}~\bibnamefont {Wyart}},\ }\href
  {https://doi.org/10.1073/pnas.1406391111} {\bibfield  {journal} {\bibinfo
  {journal} {Proceedings of the National Academy of Sciences}\ }\textbf
  {\bibinfo {volume} {111}},\ \bibinfo {pages} {14382} (\bibinfo {year}
  {2014})}\BibitemShut {NoStop}%
\bibitem [{\citenamefont {Ferrero}\ and\ \citenamefont
  {Jagla}(2019)}]{FerreroSM2019}%
  \BibitemOpen
  \bibfield  {author} {\bibinfo {author} {\bibfnamefont {E.~E.}\ \bibnamefont
  {Ferrero}}\ and\ \bibinfo {author} {\bibfnamefont {E.~A.}\ \bibnamefont
  {Jagla}},\ }\href {https://doi.org/10.1039/C9SM01073D} {\bibfield  {journal}
  {\bibinfo  {journal} {Soft Matter}\ }\textbf {\bibinfo {volume} {15}},\
  \bibinfo {pages} {9041} (\bibinfo {year} {2019})}\BibitemShut {NoStop}%
\bibitem [{\citenamefont {Bailey}\ \emph {et~al.}(2007)\citenamefont {Bailey},
  \citenamefont {Schi\o{}tz}, \citenamefont {Lema\^{i}tre},\ and\ \citenamefont
  {Jacobsen}}]{lemaitre2007avalanche-size}%
  \BibitemOpen
  \bibfield  {author} {\bibinfo {author} {\bibfnamefont {N.~P.}\ \bibnamefont
  {Bailey}}, \bibinfo {author} {\bibfnamefont {J.}~\bibnamefont {Schi\o{}tz}},
  \bibinfo {author} {\bibfnamefont {A.}~\bibnamefont {Lema\^{i}tre}},\ and\
  \bibinfo {author} {\bibfnamefont {K.~W.}\ \bibnamefont {Jacobsen}},\ }\href
  {https://doi.org/10.1103/physrevlett.98.095501} {\bibfield  {journal}
  {\bibinfo  {journal} {Phys. Rev. Lett.}\ }\textbf {\bibinfo {volume} {98}},\
  \bibinfo {pages} {095501} (\bibinfo {year} {2007})}\BibitemShut {NoStop}%
\bibitem [{\citenamefont {Halsey}(2000)}]{halsey2000diffusion}%
  \BibitemOpen
  \bibfield  {author} {\bibinfo {author} {\bibfnamefont {T.~C.}\ \bibnamefont
  {Halsey}},\ }\href {https://doi.org/10.1063/1.1333284} {\bibfield  {journal}
  {\bibinfo  {journal} {Physics Today}\ }\textbf {\bibinfo {volume} {53}},\
  \bibinfo {pages} {36} (\bibinfo {year} {2000})}\BibitemShut {NoStop}%
\bibitem [{\citenamefont {Ghosh}\ \emph {et~al.}(2017)\citenamefont {Ghosh},
  \citenamefont {Budrikis}, \citenamefont {Chikkadi}, \citenamefont {Sellerio},
  \citenamefont {Zapperi},\ and\ \citenamefont {Schall}}]{GhoshPRL2017}%
  \BibitemOpen
  \bibfield  {author} {\bibinfo {author} {\bibfnamefont {A.}~\bibnamefont
  {Ghosh}}, \bibinfo {author} {\bibfnamefont {Z.}~\bibnamefont {Budrikis}},
  \bibinfo {author} {\bibfnamefont {V.}~\bibnamefont {Chikkadi}}, \bibinfo
  {author} {\bibfnamefont {A.~L.}\ \bibnamefont {Sellerio}}, \bibinfo {author}
  {\bibfnamefont {S.}~\bibnamefont {Zapperi}},\ and\ \bibinfo {author}
  {\bibfnamefont {P.}~\bibnamefont {Schall}},\ }\href
  {https://doi.org/10.1103/PhysRevLett.118.148001} {\bibfield  {journal}
  {\bibinfo  {journal} {Phys. Rev. Lett.}\ }\textbf {\bibinfo {volume} {118}},\
  \bibinfo {pages} {148001} (\bibinfo {year} {2017})}\BibitemShut {NoStop}%
\bibitem [{\citenamefont {Leishangthem}\ \emph {et~al.}(2017)\citenamefont
  {Leishangthem}, \citenamefont {Parmar},\ and\ \citenamefont
  {Sastry}}]{LeishangthemNC2017}%
  \BibitemOpen
  \bibfield  {author} {\bibinfo {author} {\bibfnamefont {P.}~\bibnamefont
  {Leishangthem}}, \bibinfo {author} {\bibfnamefont {A.~D.~S.}\ \bibnamefont
  {Parmar}},\ and\ \bibinfo {author} {\bibfnamefont {S.}~\bibnamefont
  {Sastry}},\ }\href {https://doi.org/10.1038/ncomms14653} {\bibfield
  {journal} {\bibinfo  {journal} {Nature Communications}\ }\textbf {\bibinfo
  {volume} {8}},\ \bibinfo {pages} {14653} (\bibinfo {year}
  {2017})}\BibitemShut {NoStop}%
\bibitem [{Note5()}]{Note5}%
  \BibitemOpen
  \bibinfo {note} {Some regions can deform many times during an avalanche and
  contribute differently to its total `size' $S$ and `duration' $T$. This is
  clear in lattice models, were $S \sim T^{d_f/z}$, with $z$ the dynamical
  exponent, but if we count the total `number of activations' $S_A$, it
  typically scales with $T$ with another exponent.}\BibitemShut {Stop}%
\bibitem [{\citenamefont {Ginot}\ \emph {et~al.}(2015)\citenamefont {Ginot},
  \citenamefont {Theurkauff}, \citenamefont {Levis}, \citenamefont {Ybert},
  \citenamefont {Bocquet}, \citenamefont {Berthier},\ and\ \citenamefont
  {Cottin-Bizonne}}]{ginot2015nonequilibrium}%
  \BibitemOpen
  \bibfield  {author} {\bibinfo {author} {\bibfnamefont {F.}~\bibnamefont
  {Ginot}}, \bibinfo {author} {\bibfnamefont {I.}~\bibnamefont {Theurkauff}},
  \bibinfo {author} {\bibfnamefont {D.}~\bibnamefont {Levis}}, \bibinfo
  {author} {\bibfnamefont {C.}~\bibnamefont {Ybert}}, \bibinfo {author}
  {\bibfnamefont {L.}~\bibnamefont {Bocquet}}, \bibinfo {author} {\bibfnamefont
  {L.}~\bibnamefont {Berthier}},\ and\ \bibinfo {author} {\bibfnamefont
  {C.}~\bibnamefont {Cottin-Bizonne}},\ }\href
  {https://doi.org/10.1103/PhysRevX.5.011004} {\bibfield  {journal} {\bibinfo
  {journal} {Phys. Rev. X}\ }\textbf {\bibinfo {volume} {5}},\ \bibinfo {pages}
  {011004} (\bibinfo {year} {2015})}\BibitemShut {NoStop}%
\bibitem [{\citenamefont {Lenne}\ and\ \citenamefont
  {Trivedi}(2022)}]{lenne2022sculpting}%
  \BibitemOpen
  \bibfield  {author} {\bibinfo {author} {\bibfnamefont {P.-F.}\ \bibnamefont
  {Lenne}}\ and\ \bibinfo {author} {\bibfnamefont {V.}~\bibnamefont
  {Trivedi}},\ }\href@noop {} {\bibfield  {journal} {\bibinfo  {journal}
  {Nature Communications}\ }\textbf {\bibinfo {volume} {13}},\ \bibinfo {pages}
  {664} (\bibinfo {year} {2022})}\BibitemShut {NoStop}%
\bibitem [{\citenamefont {Mongera}\ \emph {et~al.}(2018)\citenamefont
  {Mongera}, \citenamefont {Rowghanian}, \citenamefont {Gustafson},
  \citenamefont {Shelton}, \citenamefont {Kealhofer}, \citenamefont {Carn},
  \citenamefont {Serwane}, \citenamefont {Lucio}, \citenamefont {Giammona},\
  and\ \citenamefont {Camp{\`a}s}}]{mongera2018fluid}%
  \BibitemOpen
  \bibfield  {author} {\bibinfo {author} {\bibfnamefont {A.}~\bibnamefont
  {Mongera}}, \bibinfo {author} {\bibfnamefont {P.}~\bibnamefont {Rowghanian}},
  \bibinfo {author} {\bibfnamefont {H.~J.}\ \bibnamefont {Gustafson}}, \bibinfo
  {author} {\bibfnamefont {E.}~\bibnamefont {Shelton}}, \bibinfo {author}
  {\bibfnamefont {D.~A.}\ \bibnamefont {Kealhofer}}, \bibinfo {author}
  {\bibfnamefont {E.~K.}\ \bibnamefont {Carn}}, \bibinfo {author}
  {\bibfnamefont {F.}~\bibnamefont {Serwane}}, \bibinfo {author} {\bibfnamefont
  {A.~A.}\ \bibnamefont {Lucio}}, \bibinfo {author} {\bibfnamefont
  {J.}~\bibnamefont {Giammona}},\ and\ \bibinfo {author} {\bibfnamefont
  {O.}~\bibnamefont {Camp{\`a}s}},\ }\href@noop {} {\bibfield  {journal}
  {\bibinfo  {journal} {Nature}\ }\textbf {\bibinfo {volume} {561}},\ \bibinfo
  {pages} {401} (\bibinfo {year} {2018})}\BibitemShut {NoStop}%
\bibitem [{\citenamefont {Coban}\ \emph {et~al.}(2021)\citenamefont {Coban},
  \citenamefont {Bergonzini}, \citenamefont {Zweemer},\ and\ \citenamefont
  {Danen}}]{coban2021metastasis}%
  \BibitemOpen
  \bibfield  {author} {\bibinfo {author} {\bibfnamefont {B.}~\bibnamefont
  {Coban}}, \bibinfo {author} {\bibfnamefont {C.}~\bibnamefont {Bergonzini}},
  \bibinfo {author} {\bibfnamefont {A.~J.}\ \bibnamefont {Zweemer}},\ and\
  \bibinfo {author} {\bibfnamefont {E.~H.}\ \bibnamefont {Danen}},\ }\href@noop
  {} {\bibfield  {journal} {\bibinfo  {journal} {British journal of cancer}\
  }\textbf {\bibinfo {volume} {124}},\ \bibinfo {pages} {49} (\bibinfo {year}
  {2021})}\BibitemShut {NoStop}%
\bibitem [{\citenamefont {Barbazan}\ \emph {et~al.}(2023)\citenamefont
  {Barbazan}, \citenamefont {P{\'e}rez-Gonz{\'a}lez}, \citenamefont
  {G{\'o}mez-Gonz{\'a}lez}, \citenamefont {Dedenon}, \citenamefont {Richon},
  \citenamefont {Latorre}, \citenamefont {Serra}, \citenamefont {Mariani},
  \citenamefont {Descroix}, \citenamefont {Sens} \emph
  {et~al.}}]{barbazan2023cancer}%
  \BibitemOpen
  \bibfield  {author} {\bibinfo {author} {\bibfnamefont {J.}~\bibnamefont
  {Barbazan}}, \bibinfo {author} {\bibfnamefont {C.}~\bibnamefont
  {P{\'e}rez-Gonz{\'a}lez}}, \bibinfo {author} {\bibfnamefont {M.}~\bibnamefont
  {G{\'o}mez-Gonz{\'a}lez}}, \bibinfo {author} {\bibfnamefont {M.}~\bibnamefont
  {Dedenon}}, \bibinfo {author} {\bibfnamefont {S.}~\bibnamefont {Richon}},
  \bibinfo {author} {\bibfnamefont {E.}~\bibnamefont {Latorre}}, \bibinfo
  {author} {\bibfnamefont {M.}~\bibnamefont {Serra}}, \bibinfo {author}
  {\bibfnamefont {P.}~\bibnamefont {Mariani}}, \bibinfo {author} {\bibfnamefont
  {S.}~\bibnamefont {Descroix}}, \bibinfo {author} {\bibfnamefont
  {P.}~\bibnamefont {Sens}}, \emph {et~al.},\ }\href@noop {} {\bibfield
  {journal} {\bibinfo  {journal} {Nature Communications}\ }\textbf {\bibinfo
  {volume} {14}},\ \bibinfo {pages} {6966} (\bibinfo {year}
  {2023})}\BibitemShut {NoStop}%
\bibitem [{\citenamefont {Zubiarrain-Laserna}\ \emph
  {et~al.}(2024)\citenamefont {Zubiarrain-Laserna}, \citenamefont
  {Mart{\'\i}nez-Moreno}, \citenamefont {de~Andr{\'e}s}, \citenamefont
  {de~Lara-Pe{\~n}a}, \citenamefont {Guaresti}, \citenamefont {Zaldua},
  \citenamefont {Jim{\'e}nez},\ and\ \citenamefont
  {Marchal}}]{zubiarrain2024beyond}%
  \BibitemOpen
  \bibfield  {author} {\bibinfo {author} {\bibfnamefont {A.}~\bibnamefont
  {Zubiarrain-Laserna}}, \bibinfo {author} {\bibfnamefont {D.}~\bibnamefont
  {Mart{\'\i}nez-Moreno}}, \bibinfo {author} {\bibfnamefont {J.~L.}\
  \bibnamefont {de~Andr{\'e}s}}, \bibinfo {author} {\bibfnamefont
  {L.}~\bibnamefont {de~Lara-Pe{\~n}a}}, \bibinfo {author} {\bibfnamefont
  {O.}~\bibnamefont {Guaresti}}, \bibinfo {author} {\bibfnamefont {A.~M.}\
  \bibnamefont {Zaldua}}, \bibinfo {author} {\bibfnamefont {G.}~\bibnamefont
  {Jim{\'e}nez}},\ and\ \bibinfo {author} {\bibfnamefont {J.~A.}\ \bibnamefont
  {Marchal}},\ }\href@noop {} {\bibfield  {journal} {\bibinfo  {journal}
  {Biofabrication}\ }\textbf {\bibinfo {volume} {16}},\ \bibinfo {pages}
  {042002} (\bibinfo {year} {2024})}\BibitemShut {NoStop}%
\bibitem [{\citenamefont {Gu}\ \emph {et~al.}(2025)\citenamefont {Gu},
  \citenamefont {Guiselin}, \citenamefont {Bain}, \citenamefont {Zuriguel},\
  and\ \citenamefont {Bartolo}}]{GuNature2025}%
  \BibitemOpen
  \bibfield  {author} {\bibinfo {author} {\bibfnamefont {F.}~\bibnamefont
  {Gu}}, \bibinfo {author} {\bibfnamefont {B.}~\bibnamefont {Guiselin}},
  \bibinfo {author} {\bibfnamefont {N.}~\bibnamefont {Bain}}, \bibinfo {author}
  {\bibfnamefont {I.}~\bibnamefont {Zuriguel}},\ and\ \bibinfo {author}
  {\bibfnamefont {D.}~\bibnamefont {Bartolo}},\ }\href
  {https://doi.org/10.1038/s41586-024-08514-6} {\bibfield  {journal} {\bibinfo
  {journal} {Nature}\ }\textbf {\bibinfo {volume} {638}},\ \bibinfo {pages}
  {112} (\bibinfo {year} {2025})}\BibitemShut {NoStop}%
\bibitem [{\citenamefont {Voigtländer}\ \emph {et~al.}(2024)\citenamefont
  {Voigtländer}, \citenamefont {Houssais}, \citenamefont {Bacik},
  \citenamefont {Bourg}, \citenamefont {Burton}, \citenamefont {Daniels},
  \citenamefont {Datta}, \citenamefont {Del~Gado}, \citenamefont {Deshpande},
  \citenamefont {Devauchelle}, \citenamefont {Ferdowsi}, \citenamefont {Glade},
  \citenamefont {Goehring}, \citenamefont {Hewitt}, \citenamefont {Jerolmack},
  \citenamefont {Juanes}, \citenamefont {Kudrolli}, \citenamefont {Lai},
  \citenamefont {Li}, \citenamefont {Masteller}, \citenamefont {Nissanka},
  \citenamefont {Rubin}, \citenamefont {Stone}, \citenamefont {Suckale},
  \citenamefont {Vriend}, \citenamefont {Wettlaufer},\ and\ \citenamefont
  {Yang}}]{VoiglanderSM2024}%
  \BibitemOpen
  \bibfield  {author} {\bibinfo {author} {\bibfnamefont {A.}~\bibnamefont
  {Voigtländer}}, \bibinfo {author} {\bibfnamefont {M.}~\bibnamefont
  {Houssais}}, \bibinfo {author} {\bibfnamefont {K.~A.}\ \bibnamefont {Bacik}},
  \bibinfo {author} {\bibfnamefont {I.~C.}\ \bibnamefont {Bourg}}, \bibinfo
  {author} {\bibfnamefont {J.~C.}\ \bibnamefont {Burton}}, \bibinfo {author}
  {\bibfnamefont {K.~E.}\ \bibnamefont {Daniels}}, \bibinfo {author}
  {\bibfnamefont {S.~S.}\ \bibnamefont {Datta}}, \bibinfo {author}
  {\bibfnamefont {E.}~\bibnamefont {Del~Gado}}, \bibinfo {author}
  {\bibfnamefont {N.~S.}\ \bibnamefont {Deshpande}}, \bibinfo {author}
  {\bibfnamefont {O.}~\bibnamefont {Devauchelle}}, \bibinfo {author}
  {\bibfnamefont {B.}~\bibnamefont {Ferdowsi}}, \bibinfo {author}
  {\bibfnamefont {R.}~\bibnamefont {Glade}}, \bibinfo {author} {\bibfnamefont
  {L.}~\bibnamefont {Goehring}}, \bibinfo {author} {\bibfnamefont {I.~J.}\
  \bibnamefont {Hewitt}}, \bibinfo {author} {\bibfnamefont {D.}~\bibnamefont
  {Jerolmack}}, \bibinfo {author} {\bibfnamefont {R.}~\bibnamefont {Juanes}},
  \bibinfo {author} {\bibfnamefont {A.}~\bibnamefont {Kudrolli}}, \bibinfo
  {author} {\bibfnamefont {C.-Y.}\ \bibnamefont {Lai}}, \bibinfo {author}
  {\bibfnamefont {W.}~\bibnamefont {Li}}, \bibinfo {author} {\bibfnamefont
  {C.}~\bibnamefont {Masteller}}, \bibinfo {author} {\bibfnamefont
  {K.}~\bibnamefont {Nissanka}}, \bibinfo {author} {\bibfnamefont {A.~M.}\
  \bibnamefont {Rubin}}, \bibinfo {author} {\bibfnamefont {H.~A.}\ \bibnamefont
  {Stone}}, \bibinfo {author} {\bibfnamefont {J.}~\bibnamefont {Suckale}},
  \bibinfo {author} {\bibfnamefont {N.~M.}\ \bibnamefont {Vriend}}, \bibinfo
  {author} {\bibfnamefont {J.~S.}\ \bibnamefont {Wettlaufer}},\ and\ \bibinfo
  {author} {\bibfnamefont {J.~Q.}\ \bibnamefont {Yang}},\ }\href
  {https://doi.org/10.1039/D4SM00391H} {\bibfield  {journal} {\bibinfo
  {journal} {Soft Matter}\ }\textbf {\bibinfo {volume} {20}},\ \bibinfo {pages}
  {5859} (\bibinfo {year} {2024})}\BibitemShut {NoStop}%
\bibitem [{\citenamefont {Fily}\ and\ \citenamefont
  {Marchetti}(2012)}]{fily2012athermal}%
  \BibitemOpen
  \bibfield  {author} {\bibinfo {author} {\bibfnamefont {Y.}~\bibnamefont
  {Fily}}\ and\ \bibinfo {author} {\bibfnamefont {M.~C.}\ \bibnamefont
  {Marchetti}},\ }\href {https://doi.org/10.1103/PhysRevLett.108.235702}
  {\bibfield  {journal} {\bibinfo  {journal} {Phys. Rev. Lett.}\ }\textbf
  {\bibinfo {volume} {108}},\ \bibinfo {pages} {235702} (\bibinfo {year}
  {2012})}\BibitemShut {NoStop}%
\bibitem [{\citenamefont {Irving}\ and\ \citenamefont
  {Kirkwood}(1950)}]{irving1950statistical-mechanical-hydrodynamics}%
  \BibitemOpen
  \bibfield  {author} {\bibinfo {author} {\bibfnamefont {J.~H.}\ \bibnamefont
  {Irving}}\ and\ \bibinfo {author} {\bibfnamefont {J.~G.}\ \bibnamefont
  {Kirkwood}},\ }\href {https://doi.org/10.1063/1.1747782} {\bibfield
  {journal} {\bibinfo  {journal} {J. Chem. Phys.}\ }\textbf {\bibinfo {volume}
  {18}},\ \bibinfo {pages} {817–829} (\bibinfo {year} {1950})}\BibitemShut
  {NoStop}%
\end{thebibliography}%
\end{document}